\documentclass[amsmath,amssymb,11pt]{article}
\usepackage{jheppub2}
\pdfoutput=1

\usepackage{amsmath,amssymb,amsthm,mathrsfs,bbm,bm}
\usepackage{latexsym,amscd,amsbsy,amsfonts,dsfont}
\usepackage{multirow}
\usepackage{graphicx}
\usepackage{xcolor}
\usepackage{caption}

\usepackage{chngpage} 

\usepackage[normalem]{ulem} 

\newcommand{\beq}{\begin{equation}}
\newcommand{\eeq}{\end{equation}}
\newcommand{\beqa}{\begin{eqnarray}}
\newcommand{\eeqa}{\end{eqnarray}}
\newcommand{\bea}{\begin{eqnarray}}
\newcommand{\eea}{\end{eqnarray}}

\newcommand*{\affmark}[1][*]{\textsuperscript{#1}} 

\numberwithin{equation}{section}  
\allowdisplaybreaks

\title{String Theory in a Pinch: Resolving the Gregory-Laflamme Singularity}
\subheader{\begin{flushright}
\texttt{CPHT-RR082.112024}
\end{flushright}}

\author{Roberto Emparan,\affmark[1,2]}
\emailAdd{emparan@ub.edu}
\author{Mikel~Sanchez-Garitaonandia\affmark[3]}
\emailAdd{mikel.sanchez@polytechnique.edu}
\author{and Marija Tomašević\affmark[4]}
\emailAdd{m.tomasevic@uva.nl}

\affiliation{
\affmark[1]Institució Catalana de Recerca i Estudis Avançats (ICREA)
 Passeig Lluis Companys, 23, 08010 Barcelona, Spain\\
\affmark[2]Departament de Física Quàntica i Astrofísica and
  Institut de Ciències del Cosmos,
 Universitat de Barcelona, 08028 Barcelona, Spain\\
\affmark[3]CPHT, CNRS, \'Ecole polytechnique, Institut Polytechnique de Paris, 91120 Palaiseau, France\\
\affmark[4]Institute for Theoretical Physics, University of Amsterdam, Science Park 904, 1090 GL Amsterdam,
The Netherlands}

\abstract{Thin enough black strings are unstable to growing ripples along their length, eventually pinching and forming a naked singularity on the horizon. We investigate how string theory can resolve this singularity. 
First, we study the string-scale version of the static non-uniform black strings that branch off at the instability threshold: ``string-ball strings'', which are linearly extended, self-gravitating configurations of string balls obtained in the Horowitz-Polchinski (HP) approach to near-Hagedorn string states. 
We construct non-uniform HP strings in spatial dimensions $d\leq 6$ and show that, as the inhomogeneity increases, they approach localized HP balls.
We also examine the thermodynamic properties of the different phases in the canonical and microcanonical ensembles. We find that, for a sufficiently small mass, the uniform HP string will be stable and not evolve into a non-uniform or localized configuration. 
Building on these results and independent evidence from the evolution of the black string instability with $\alpha'$ corrections, we propose that, at least in $d=4,5$, string theory slows and eventually halts the pinching evolution at a classically stable stringy neck. In $d\geq 6$ this transition is likely to occur into a puffed-up string ball.  The system then enters a slower phase in which the neck gradually evaporates into radiation. 
We discuss this scenario as a framework for understanding how string theory resolves the formation of naked singularities.}

\begin{document}

\maketitle


\section{Introduction}
\label{sec:intro}

The Gregory-Laflamme (GL) instability is a fundamental and ubiquitous phenomenon in the physics of higher-dimensional black holes, where horizons that extend much more in some directions than in others become unstable to growing ripples along the longer directions \cite{Gregory:1993vy,Gregory:1994bj}. 

In this article we are motivated by the close relationship between black holes and the highly excited states of fundamental strings---made concrete in the black hole/string correspondence of \cite{Susskind:1993ws,Horowitz:1996nw,Horowitz:1997jc}---to explore this phenomenon within string theory. 
The deep non-linear growth of the GL instability has been shown to lead to the appearance of naked singularities on the horizon, which makes this system a natural setting to investigate how string theory may resolve these singularities.

\paragraph{Black-string naked singularities and stringy effects.} Considering the illustrative case of a thin, long, neutral black string, we can find naked curvature singularities in two different circumstances: 
\begin{enumerate}
    \item Late time evolution of generic initial perturbations of the unstable black string, which grow until the horizon pinches off at a thin neck, thus violating cosmic censorship \cite{Lehner:2010pn,Figueras:2022zkg}. 
    \item Evolution along the space of static solutions of increasingly non-uniform black strings. This branch of solutions arises from the static zero-mode at the threshold of the instability \cite{Wiseman:2002zc,Kleihaus:2006ee,Sorkin:2006wp,Harmark:2007md,Figueras:2012xj,Kalisch:2016fkm,Kalisch:2017bin,Cardona:2018shd}. When followed non-linearly, the non-uniformity of the black strings grows until a solution is reached where the horizon pinches at a conifold where the curvature diverges \cite{Kol:2002xz,Kol:2003ja}. 
\end{enumerate}
The two scenarios, time-dependent and static, are interesting in themselves and have been thoroughly analyzed, but time evolution can be computationally very costly while the construction of static solutions is much easier. Thus, the latter has often been regarded as the simpler means of obtaining information about black string naked singularities, even though their detailed structure can be quite different in the two cases.

\begin{figure}[t]
\begin{center}
    \includegraphics[width=0.5\textwidth]{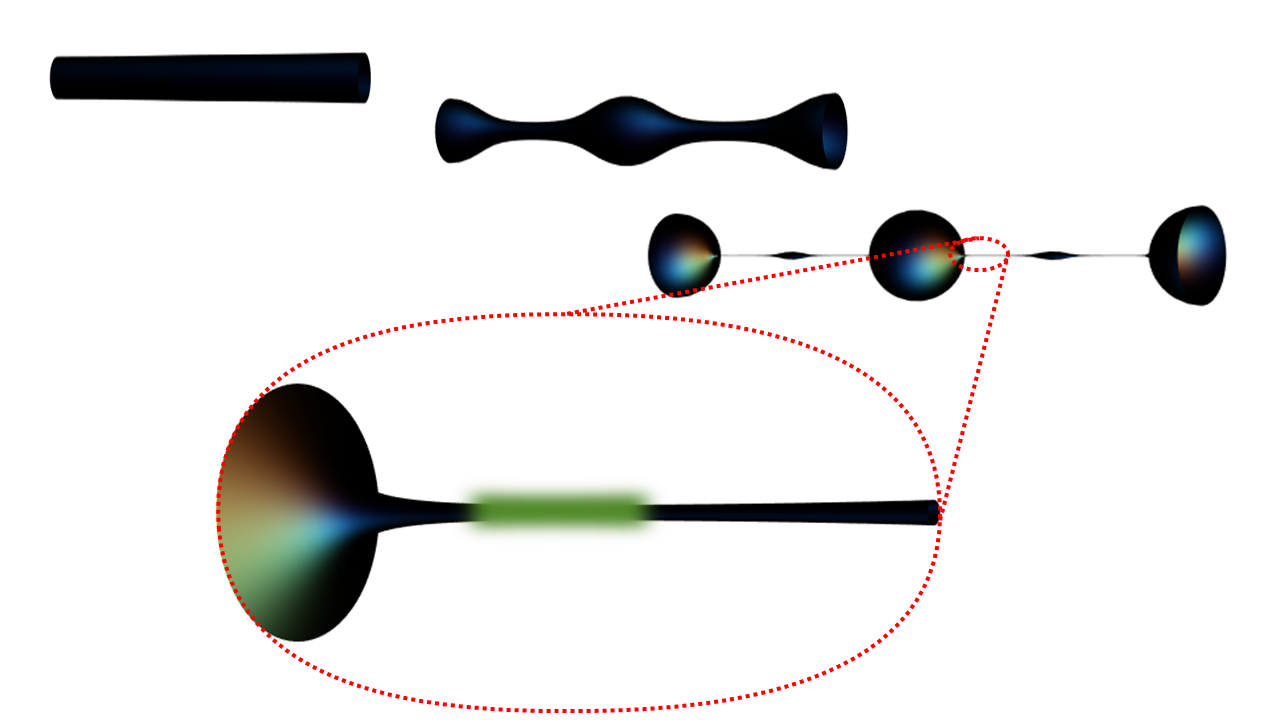}
    \quad
    \includegraphics[width=0.46\textwidth]{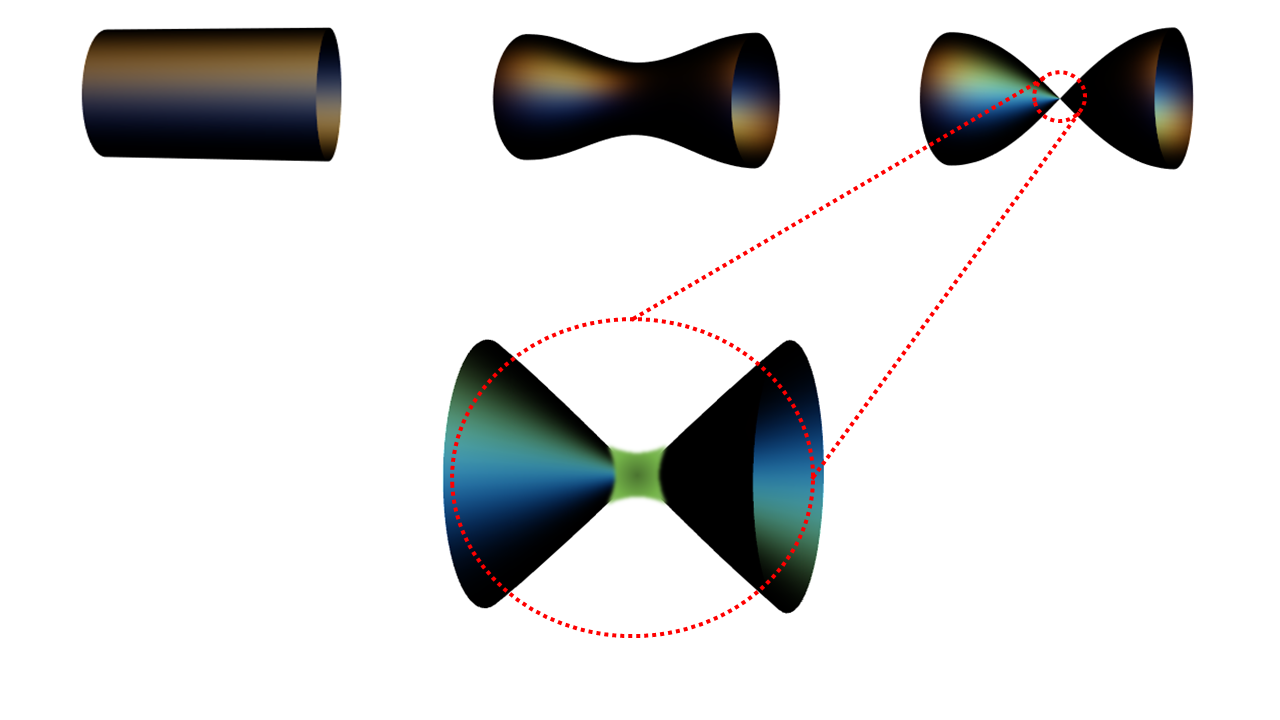}
\end{center}
\caption{\small Left: Dynamical formation of a pinch during the evolution of the black string instability. When the curvature at the pinching neck reaches the string scale, it is expected to transition into a stringy regime (fuzzy green). In this article, we describe this stringy phase and argue that the pinching neck halts and eventually evaporates into radiation at the Hagedorn temperature, completing the breakup of the black string. More complex scenarios might introduce transient variations. Right: Pinching along the solution space of static non-uniform black strings. While we expect string theory to smooth out the singular conical pinch, our understanding of this resolution remains incomplete.} 
  \label{fig:1}
\end{figure}

The black hole/string correspondence suggests a natural way of dealing with these singularities. A black hole whose horizon curvature is near the string scale is expected to morph into a highly excited string ball. In the two scenarios above, when the curvature along the horizon reaches the string scale, a transition of roughly this kind should prevent the appearance of naked singularities. In the time-evolving situation, string theory should control the further evolution of the string ball and provide a plausible mechanism by which the black string is severed into separate horizons (see Fig.~\ref{fig:1}~left).
At first glance, it would seem very difficult to make precise this picture of the evolution, due to the significant time-dependence involved, but, remarkably, there are good indications suggesting that the evolution slows down as the string scale is approached, making it more amenable to study (see Sec.~\ref{sec:notnear}).

On the other hand, the curvature pinch on the highly non-uniform static black strings seems to offer a loosely similar picture (see Fig.~\ref{fig:1}~right). Ideally, string theory would resolve the singular cone geometry that locally models the pinch. However, it seems difficult to tackle this problem with current methods (see Sec.~\ref{sec:statpinch}).

\paragraph{Stringy Gregory-Laflamme.} We will show that progress in understanding stringy effects in the GL setup can be made if 
we consider that the black string reaches string-scale thickness along a portion of its horizon. Then we can regard that part of the black string as having transitioned into a `string-ball string', that is, an object made of highly excited string that extends along a linear direction. We may envision it as the result of starting with a black string and lowering the gravitational coupling until the curvature of the horizon reaches the string scale. Although this system may not exactly match the ones mentioned above, we will extract from it useful lessons about the stringy aspects of the GL instability, as well as about the broader connections between black holes and self-gravitating fundamental strings. In particular, we will be able to investigate in detail the properties of the stringy counterparts of non-uniform black strings and learn about their stability.

\paragraph{HP strings.} Our approach is based on the formalism for self-gravitating string states near the Hagedorn temperature developed by Horowitz and Polchinski (HP) in \cite{Horowitz:1997jc}. Highly excited string states are described in terms of a mean field---the thermal winding scalar---that couples to the Newtonian potential of weak gravitational fields. Under its own gravity, the thermal scalar condenses to form `string balls' that solve the same equations as static Newtonian boson stars \cite{Friedberg:1986tp,Jones:1995wb}. There are two important differences with the latter: (i) the thermal scalar is intrinsically a field in Euclidean space and can only describe finite-temperature equilibrium configurations---static or stationary, but not time-dependent ones. As mentioned, this may pose less of a problem for our purposes than it initially appears and, in any case, serves our goal of constructing and analyzing non-uniform HP solutions.
(ii) The thermal scalar captures the thermodynamic properties of the string ball, in particular its entropy; 
boson stars are, instead, zero-temperature and zero-entropy coherent condensates.

Ref.~\cite{Chen:2021dsw} argued that HP balls have a negative Euclidean mode analogous to that of the Euclidean Schwarzschild solution. It has long been known that this mode can be alternatively identified as the static zero-mode at the threshold of the GL instability \cite{Gregory:1987nb,Reall:2001ag}, and this is the case for black strings and HP strings alike. This suggests that HP strings may exhibit GL instabilities. While the HP formalism can not directly uncover unstable modes, it motivates us to further investigate this possibility using all available tools.

We will explicitly obtain the zero modes for HP strings and find a notable agreement between their wavelength values and those of black string zero modes. As we will discuss shortly, uniform HP strings below the Hagedorn temperature exist only in spatial dimensions $d=4,5,6$ (spacetime dimension $D = d+1=5,6,7$). For these cases, the numerical results for the zero-mode wavelength, in 
units of the thickness of the string, are
\begin{equation}\label{eq:Lstar}
    \textsf{L}_*=
    \begin{cases}
        7.96,\ 4.84,\ 2.77 \qquad \textrm{HP strings}\,,\\
        7.17,\ 4.95,\ 3.98 \qquad \textrm{black strings}\,.
    \end{cases} 
\end{equation}
The closeness between the two sets of values is noteworthy, given the markedly different form of the equations governing each system.

Next, we move beyond this perturbative analysis to construct the branch of non-uniform HP strings in a Kaluza-Klein circle of length $L$, continuing until the non-uniformity is so large that the solutions are better viewed as HP balls localized within the Kaluza-Klein circle. Here, we observe one example of a stringy resolution of GL singularities: unlike the transition from black strings to localized black holes, which involves a singular change in horizon topology, HP strings and localized balls are smoothly connected, with no discontinuity.

In contrast to the similarity in \eqref{eq:Lstar} for the linearized perturbations, the non-linear regime of HP solutions in a KK circle exhibits a phase structure that is qualitatively distinct from that of black strings. The properties of HP balls show a marked dependence on the spatial dimension $d$. In particular, at any temperature below the Hagedorn limit they exist only for $d=3,4,5$ because, in higher dimensions, the gravitational potential that binds them becomes too short-ranged to counterbalance the outward pressure of the condensate \cite{Horowitz:1997jc}.\footnote{In $d=6$ there are solutions of the HP theory at the Hagedorn temperature $T_H$ \cite{Balthazar:2022szl}, but when including corrections the localized balls exist only for $T<T_H$ \cite{Balthazar:2022hno}.} As a result, uniform HP strings are limited to $d=4,5,6$. 
Non-uniform HP configurations reveal this tension: as non-uniformity increases, the solutions in $d=4$ and $5$ interpolate between a uniform string and a localized ball. However, in $d=6$, the condensate becomes less localized and spreads out, since no solution of the HP theory with exponential fall-off exists. In this case, it seems necessary to consider corrections to the effective HP theory \cite{Balthazar:2022hno}, and a phase of nearly free string balls that are very large and weakly gravitationally bound.

\paragraph{From thermodynamics to strings at a pinch.} By examining their thermodynamic properties, we find that the relative dominance of the non-uniform and uniform solutions differs between $d=4,5$ and $d=6$, as well as between the microcanonical and canonical ensembles: 
\begin{enumerate}
    \item[(a)] In the microcanonical ensemble of fixed mass and fixed circle length, uniform self-gravitating HP strings are preferred in $d=4,5$, while in $d=6$ the preferred phases are free strings. Non-uniform HP strings, including localized HP balls, are never the dominant microcanonical phase.
    \item[(b)] In the canonical ensemble of fixed temperature and circle length, non-uniform HP solutions always dominate in $d=4,5$. In $d=6$ uniform HP strings are the dominant phase, except very close to the Hagedorn temperature, where the description is in terms of (nearly) free strings.
\end{enumerate}
Most of these features are hinted at by comparing the analytically known properties of uniform strings and localized balls, as we will do in the next section. However, this comparison is heuristic. To properly analyze the thermodynamic competition between uniform and non-uniform phases, we will explicitly construct the latter. Computing their thermodynamic properties then allows us to draw the conclusions outlined above.

With these results in hand and incorporating the properties of black hole phases, we will revisit the questions posed at the beginning of the article and propose how string theory may resolve the pinching singularities in the evolution of unstable black strings. 
We will argue that the results of the HP formalism, although inherently limited to static configurations, can be meaningfully applied in the final stages of the GL instability. This is supported by recent evidence (\cite{Figueras:2025}, unpublished) that the pinching evolution of black strings decelerates as $\alpha'$ corrections to the gravitational action become significant and the black hole/string changeover is near. Therefore, the static HP solutions we have constructed can be the crucial transitional configuration leading to the non-singular endpoint of the GL instability.

\smallskip

The remainder of the article is organized as follows: The next section provides an elementary, heuristic study of the thermodynamic competition between uniform HP strings and localized balls, concluding, in particular, that the latter never dominate the microcanonical ensemble. This sets the stage for our subsequent detailed analysis, which begins in Section~\ref{sec:hpstrings} with the introduction of the HP framework and the construction of its solutions. In Section~\ref{sec:entropy_results} we examine the thermodynamic phases and their stability in both the microcanonical and canonical ensembles. We also add black hole phases to the diagram, since they are relevant for masses above the correspondence transition. These results are then employed in Section~\ref{sec:resolution} to argue for a stringy resolution of the GL singularities. Finally, in Section~\ref{sec:conclusion}, we conclude with brief remarks on open directions for future work.

Readers primarily interested in our proposal for resolving the GL singularity may wish to begin with Section~\ref{sec:basics}, and then skip to the final paragraphs of Section~\ref{sec:stabcrit}, which summarize the key information needed for the arguments presented in Section~\ref{sec:resolution}.

\bigskip 

\noindent\emph{Note:} While we were writing up our results, Ref.~\cite{Chu:2024ggi} appeared in arXiv, which performs an analysis of non-uniform HP solutions in a perturbative expansion for small non-uniformity and then studies the phase diagram. The results from both studies agree where they overlap.

\section{Thermodynamic comparison between HP strings and balls}\label{sec:basics}

In \cite{Gregory:1993vy,Gregory:1994bj} a simple argument was given for why black strings might be expected to evolve into localized black holes. The entropy of a black hole in $D=d+1$ dimensions takes the form
\begin{align}\label{SMbh}
    S_{BH}=c_d\, M^{\frac{d-1}{d-2}}\,
\end{align}
(the value of $c_d$ is not needed for the argument). Now we imagine that this black hole is localized in a circle of length $L$, and neglect the finite-size distortions that would modify the entropy formula.

Next, we consider a black string in the same number of dimensions, obtained by uniformly translating a $(d-1)$-dimensional black hole. The formula \eqref{SMbh} applies after replacing $d\to d-1$ and scaling $S\to S/L$ and $M\to M/L$, so the entropy is
\begin{align}\label{SMst}
    S_{BS}=c_{d-1} M^{\frac{d-2}{d-3}} L^{-\frac1{d-3}}\,.
\end{align}
It is now clear that if we compare a black hole and a black string of the same mass, then for $L$ sufficiently larger than $M$, it will be entropically favorable for a black string to transition into a localized black hole. Note that the argument indicates a thermodynamic instability but it does not imply a dynamic instability, namely, there may not be any classical evolution connecting the two systems. If the black string is unstable, it may happen that the classical evolution takes it to a stable non-uniform black string, and not all the way down to a fully localized black hole, which would involve a singular transition. There is strong evidence (summarized in \cite{Emparan:2018bmi}) that the actual evolution and endpoint of the instability depends on the number of dimensions. 

Let us now apply the same simple argument to HP balls and strings using 
the results of \cite{Chen:2021dsw} for HP balls, which we will rederive below, in the form
\begin{align}\label{SMHPb}
    S_{b} = \beta_H M + \textsf{g}_d \frac{d-4}{d-6}M^{\frac{d-6}{d-4}}\,.
\end{align}
The factor $\textsf{g}_d$ is positive in any $d$ and multiplies the self-interaction corrections to the entropy $S=\beta_H M$ of the free string, at a temperature slightly below the Hagedorn temperature $T_H=\beta_H^{-1}\sim 1/\sqrt{\alpha'}$. In $d=4$ no correction is obtained within the approximations of the model, which gives ball solutions with the same mass independently of the temperature. We assume again that this equation provides a good estimation of the entropy of an HP ball localized in a long enough circle $L$.

For a uniform HP string, the same reasoning as above gives
\begin{align}\label{SMHPs}
    S_{s} = \beta_H M + \textsf{g}_{d-1} \frac{d-5}{d-7}M^{\frac{d-7}{d-5}}L^{\frac2{d-5}}\,.
\end{align}
Recall that HP strings exist only in $d=4,5,6$. Again, in $d=5$ there is no correction to the entropy within this model.

We can now compare the entropies of an HP ball and an HP string of the same mass. We observe that in $d=4$ the correction for a string is $\sim +1/L^2$ while the ball receives no correction. So, for any non-zero length $L$, the string is always more entropic than the ball and therefore will be thermodynamically preferred.
In $d=5$ the ball entropy is corrected by a negative term, while the string receives no correction. So, again, the uniform HP string is thermodynamically favored.

This suggests that, at least in $d=4,5$, \emph{long, uniform HP strings are unlikely to break up into rounded string balls.} This conclusion is in stark contrast with the behavior of black objects, which are always expected to break up if they are sufficiently long. 
To be clear, the analysis will not be complete without considering both the black hole and HP solutions, including the possible correspondence transitions between them. For now, we focus on comparing only the HP phases and will defer the more comprehensive discussion to Sec.~\ref{subsec:allphases}.

In $d=6$ the situation is more complicated. The string entropy is reduced by $\sim -L^2$ and the approximations break down for large enough $L$. But in this dimension the string cannot evolve into a localized HP ball, since the latter does not exist at any temperature below the Hagedorn limit. Instead, very large, puffed-up balls have been found in $d=6$ at the Hagedorn temperature and with entropy $S\simeq \beta_H M$ \cite{Balthazar:2022szl}, but then  Ref.~\cite{Balthazar:2022hno} argues that corrections to the HP theory change this conclusion and localized balls exist only for $T<T_H$. The dominant phase is likely to be a very large, almost-free string ball with Hagedorn entropy. The same conclusion is expected in $d>6$.

The canonical ensemble of phases at fixed temperature is rarely discussed for black objects in this context, since it gives the same conclusions as the microcanonical analysis. Interestingly, for HP strings and balls the situation turns out to be reversed in the two ensembles. 
The free energy for temperatures near the Hagedorn limit can be approximated by
\begin{equation}\label{eq:freeF}
    F=M-TS\simeq \left(1-\frac{T}{T_H}\right) M - T_H \delta S\,,
\end{equation}
where $\delta S= S-\beta_H M$ is the self-gravitation correction to the entropy. For the HP ball, this gives
\begin{equation}\label{eq:Fball}
    F_b=\frac{2}{6-d}\textsf{g}_d^\frac{d-4}{2}\left(1-\frac{T}{T_H}\right)^\frac{6-d}{2}\,,
\end{equation}
and for the uniform HP string,
\begin{equation}\label{eq:Fustring}
    F_s=L\frac{2}{7-d}\textsf{g}_{d-1}^\frac{d-5}{2}\left(1-\frac{T}{T_H}\right)^\frac{7-d}{2}\,.
\end{equation}
The comparison is simple. Given that the dimension factors are positive, when $L$ is large enough the HP ball will have lower free energy and thus dominate in $d=4,5$. For $d=6$, the ball solution is not well defined at any $T<T_H$ (without corrections to the HP theory). The apparent blow-up of $F_b$ seems to indicate that non-uniformity increases the free energy.

In the next two sections, we will confirm these preliminary conclusions with a detailed study of non-uniform HP strings and their thermodynamics.

\section{Self-gravitating strings}
\label{sec:hpstrings}

The highly excited states of string theory near the Hagedorn temperature 
can be collectively described, in the Euclidean time formalism, by an effective mean-field $\chi$. This is the winding mode of the string around the Euclidean time circle, which becomes almost massless when its length is $\beta\simeq \beta_H$ \cite{Polchinski:1985zf,Atick:1988si}. Being light, this field must be added to the effective action of string theory at low energies, which also contains the graviton and dilaton. The coupling between the latter and $\chi$ allows to describe self-gravitating configurations of highly-excited strings, often called string balls or string stars \cite{Horowitz:1997jc}. 

\subsection{Framework}

After integrating the Euclidean time circle, the effective action in the $d$ non-compact spatial directions becomes \cite{Horowitz:1996cj,Chen:2021dsw}
\begin{equation}\label{eq:Sb}
    I_d = \frac{1}{16\pi G_N}\int d^dx\sqrt{g}\,e^{-2\phi_d}\left(-\mathcal{R}-4(\nabla\phi_d)^2+(\nabla \varphi)^2+\vert\nabla\chi\vert^2+m(\varphi)^2\vert\chi\vert^2+\dots\right)\,.
\end{equation}
Here $\phi_d$ and $g_{ab}$ are the $d$-dimensional dilaton and spatial metric. The length of the Euclidean time circle is $\beta e^\varphi$, so $\varphi$ is the gravitational potential in $d$ dimensions. The mass of the thermal scalar $\chi$
\begin{equation}\label{eq:m2}
    m(\varphi)^2 = m^2_{\infty} + \frac{\kappa}{\alpha '}\varphi + \mathcal{O}(\varphi^2), \qquad m_{\infty}^2 = \frac{\kappa}{\alpha '}\frac{\beta-\beta_H}{\beta_H}\,,
\end{equation}
depends on $\varphi$ and takes the value $m^2_\infty$ at large distances where $\varphi\to 0$. The constant $\kappa$ depends on the specific type of string (heterotic, type II or bosonic) and we will not need its precise value. The dots denote higher-order terms and we will work in regimes where they are negligible (see e.g., \cite{Brustein:2021ifl,Balthazar:2022hno}). Although in general $\chi$ is complex, all our configurations will be neutral under the string two-form potential $B_{ab}$ and we can take $\chi$ to be real and positive.

Very close to the Hagedorn temperature, when the winding scalar is very light, $m_{\infty}^2\ll \kappa/\alpha'\sim M_s^2$, the dominant interaction is the one between $\varphi$ and $\chi$. The dilaton $\phi_d$ and the spatial metric $g_{ab}$ can consistently remain fixed and the field equations for $\varphi$ and $\chi$ are
\begin{equation}
    \begin{aligned}
        \nabla ^ 2\chi -\left(m_{\infty}^2+\frac{\kappa}{\alpha '}\varphi\right)\chi & =0,\\
        \nabla ^ 2 \varphi -\frac{1}{2}\frac{\kappa}{\alpha '}\chi^2 & = 0.
    \end{aligned}
\end{equation}
Redefining the spatial coordinates
\begin{equation}
    x \to \sqrt{\frac{\alpha'}{\kappa}}x
\end{equation}
to measure all lengths in string units, the equations become\footnote{The condensate is often rescaled $\chi\to\sqrt{2}\chi$ and the potential shifted $\varphi+\Delta_\beta\to \varphi$.}
\begin{equation}
    \begin{aligned}
        \nabla^2\chi -\left(\Delta_\beta+\varphi\right)\chi & =0,\\
        \nabla^2\varphi -\frac12\chi^2 & = 0\,.
    \end{aligned}
    \label{eq:eigensystem}
\end{equation}

Since we are interested in the thermodynamics of this system, we have changed the notation from $m^2_\infty$ to
\begin{equation}
    \Delta_\beta \equiv\frac{\beta-\beta_H}{\beta_H}=\frac{T_H}{T}-1\,, \label{deltabeta}
\end{equation}
to emphasize the meaning of this parameter as controlling the temperature of the system relative to its value at the Hagedorn point, always from below: $T<T_H$, so $\Delta_\beta>0$. 

The equations \eqref{eq:eigensystem} are the same as for a non-relativistic boson star where a boson condensate $\chi$ is coupled to the Newtonian potential $\varphi$ \cite{Friedberg:1986tp,Jones:1995wb}.\footnote{In \cite{Albertini:2024hwi}, a self-interacting scalar field model without gravity is shown to exhibit similar properties.} One important difference, though, is that the HP formalism is purely Euclidean and cannot be used to study time-dependent fluctuations (e.g., quasinormal modes) of the string ball. 

\paragraph{Validity.} The applicability of this theory is limited by two factors. First, in order to neglect the corrections to the effective theory, the condensate mass and the field amplitude must be small $\Delta_\beta\,,|\chi|\ll 1$. When these become $O(1)$, higher non-linearities in the effective theory become important and the balls eventually become black holes. 

A second limitation occurs very close to the Hagedorn temperature, where quantum fluctuations can become very large \cite{Horowitz:1997jc}. These are controlled by the competition between the string coupling $g$ (which, when small, suppresses quantum effects) and the mass $\Delta_\beta$ of the condensate (whose fluctuations can be large when very light). The effect was estimated in \cite{Chen:2021dsw}, and, combined with the previous requirement, leads to the conclusion that the classical HP configurations are valid only when
\begin{equation}\label{eq:qlimit}
   g^\frac{4}{6-d}\lesssim \Delta_\beta \ll 1\,.
\end{equation}
When closer to the Hagedorn limit than the lower bound the classical solutions have $O(1)$ action and they are not good saddle points.  The weakness of the gravitational interaction suggests that the appropriate description of the string state is as a free string with Hagedorn entropy $S=\beta_H M$ and random-walk size $\sim \sqrt{S}$ (in string units). At the other end, when $\Delta_\beta\sim 1$ the HP balls transition into black holes. As $d=6$ is approached from below, the upper range of validity of the HP theory becomes smaller and in $d\geq 6$ corrections to the effective theory must be included \cite{Balthazar:2022hno}.

\paragraph{Scaling symmetry.}
A crucial property of the equations \eqref{eq:eigensystem} is their invariance under the rescaling
\begin{equation}\label{eq:scaling}
    (x^i,\chi,\varphi,\Delta_\beta)\to(\lambda^{-1/2} x^i,\lambda\chi,\lambda\varphi, \lambda\Delta_\beta)\,.
\end{equation}
This implies that, for any given solution to the equations, a simple rescaling produces another solution with a different amplitude, size, and temperature. At first glance, this may seem surprising, as the solutions are valid only for small field amplitudes and temperatures close to the Hagedorn limit. However, this scaling symmetry is only approximate and is broken by the corrections to the HP effective action or the quantum fluctuations discussed above.

The situation is very similar to that of black holes in gravity: in the full quantum theory, where the Planck length can be set to one, we can consider semiclassical configurations with sizes $\gg 1$ in Planck units. In this classical limit, vacuum gravity possesses a scaling symmetry that relates black holes of different sizes, but, again, this is only an approximate symmetry that is broken when Planck-scale effects become relevant. We see that gravity and strings near the Hagedorn temperature share a key similarity: they possess classical limits where the fundamental length scale appears solely as an overall factor in their respective actions and the field equations are scale-invariant. They differ, though, in that the HP theory has both lower and upper limits of validity \eqref{eq:qlimit}, whereas classical gravity only fails at very short distances.\footnote{Although there are intriguing hints of the failure of gravity at large scales.}

This scaling symmetry lies behind the form of the equation of state $S(M)$ that we used in Sec.~\ref{sec:basics}, and which we will rederive below. The scaling symmetry in neutral black holes also implies their $S(M)$ up to a factor.

Another consequence of \eqref{eq:scaling} is that, since  $x^i \sqrt{\Delta_\beta}$ is scale-invariant, the spatial extent of a string ball diverges when $\beta\to\beta_H$ and $\Delta_\beta\to 0$. Therefore, very close to the Hagedorn temperature the ball spreads out over increasingly larger distances. As we will shortly explain, $1/\sqrt{\Delta_\beta}$ measures the radius of the ball.

In the following, we will often work with scale-invariant quantities, which we denote with sans-serif fonts.

\subsection{HP balls}
\label{sec:HP-balls}

We begin by reproducing the results of \cite{Chen:2021dsw} for HP string balls. These correspond to spherically symmetric configurations where $\varphi$ and $\chi$ vanish asymptotically and the condensate is regular at the origin,
\begin{equation}\label{eq:bcs}
    \varphi(r)\,,\chi(r)\to 0\quad\text{as}~r\to\infty\,,\qquad \partial_r \chi(0)=0\,.
\end{equation}
One more condition is needed to completely specify the boundary value problem for \eqref{eq:eigensystem}. 
To this effect, notice that the scaling symmetry \eqref{eq:scaling} (which is preserved by the boundary conditions \eqref{eq:bcs}) implies that we are free to fix the overall amplitude of the condensate, e.g., by selecting an arbitrary value for $\chi(0)$.

Having chosen the amplitude of $\chi$, the boundary value problem becomes similar to an eigenvalue problem (and we will refer to it as such): 
we must find the eigenvalue $\Delta_\beta$ for which \eqref{eq:eigensystem} admit a solution with these boundary conditions.\footnote{In \cite{Chen:2021dsw} the eigenvalue problem is fixed not by selecting a value for $\chi(0)$, but instead by demanding that $\chi$ be unit-normalized.} Having the solution we can extract its mass from the asymptotic fall-off of $\varphi$,
\begin{equation}\label{asymppot}
    \varphi(r)= -\frac{8\pi}{(d-2)\omega_{d-1}}\frac{G_N M}{r^{d-2}}+O(r^{1-d})\,.
\end{equation}
By Gauss's theorem, this is proportional to the norm of $\chi$, and it is easy to verify that this $M$ agrees with the thermodynamic mass obtained from the Euclidean action \cite{Chen:2021dsw}.
Thus we obtain the temperature and mass of a solution of a given amplitude. By rescaling it, we can find the relation $\beta(M)$ for the string ball states in $d$ dimensions.

\paragraph{Solutions.} The equations for this problem can be readily solved with conventional numerical techniques. Spherical balls involve radial ODEs for which a shooting method is good enough, but since later we deal with PDEs, we resort to a relaxation method as in, e.g., \cite{Dias:2015nua}.

To this effect, we compactify the radial direction by changing to 
\begin{equation}\label{eq:rtou}
    u = \left(\frac{r}{1+r}\right)^2\,, 
\end{equation}
and then discretize the domain $u\in [0,1]$ using a Lobatto-Chebyshev grid. Regularity at the origin is automatically satisfied when using $u$, and the asymptotic boundary conditions are $\chi(1)=\varphi(1)=0$. In our calculations, we have arbitrarily chosen $\chi(0)=1$ for the reference ball. Strictly speaking, we should only consider small values of $\chi$, but we can always scale down the amplitude of the solution as much as we wish. Thus the choice above is harmless and slightly convenient.

\begin{figure}
\begin{center}
    \includegraphics[width=0.46\textwidth]{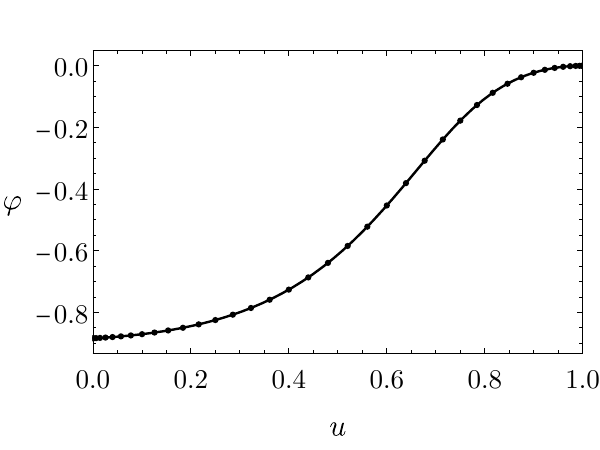}
    \quad
    \includegraphics[width=0.45\textwidth]{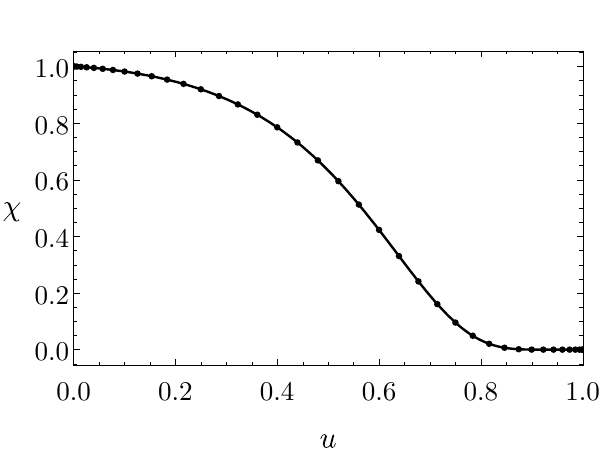}
\end{center}
\caption{\small String ball solution for $d=4$ as a function of the compactified radial coordinate $u=(r/(1+r))^2$. Left: gravitational potential $\varphi$. Right: scalar condensate $\chi$. The temperature parameter is $\Delta_\beta=0.1996\, \chi(0)$. The vertical axes are measured in units where $\chi(0)=1$.} 
  \label{fig:HP-ball}
\end{figure}

Ball solutions for $d=3,4,5$ can now be easily obtained. In Fig.~\ref{fig:HP-ball} we show the profiles we obtain for $\varphi$ and $\chi$ in $d=4$, the other two cases behaving similarly. In every $d$ there is a discrete spectrum of states, and the excited states of string balls are also easily found, but we are only interested in the lowest mass state.

\begin{figure}
    \begin{center}
        \includegraphics[width=0.47\textwidth]{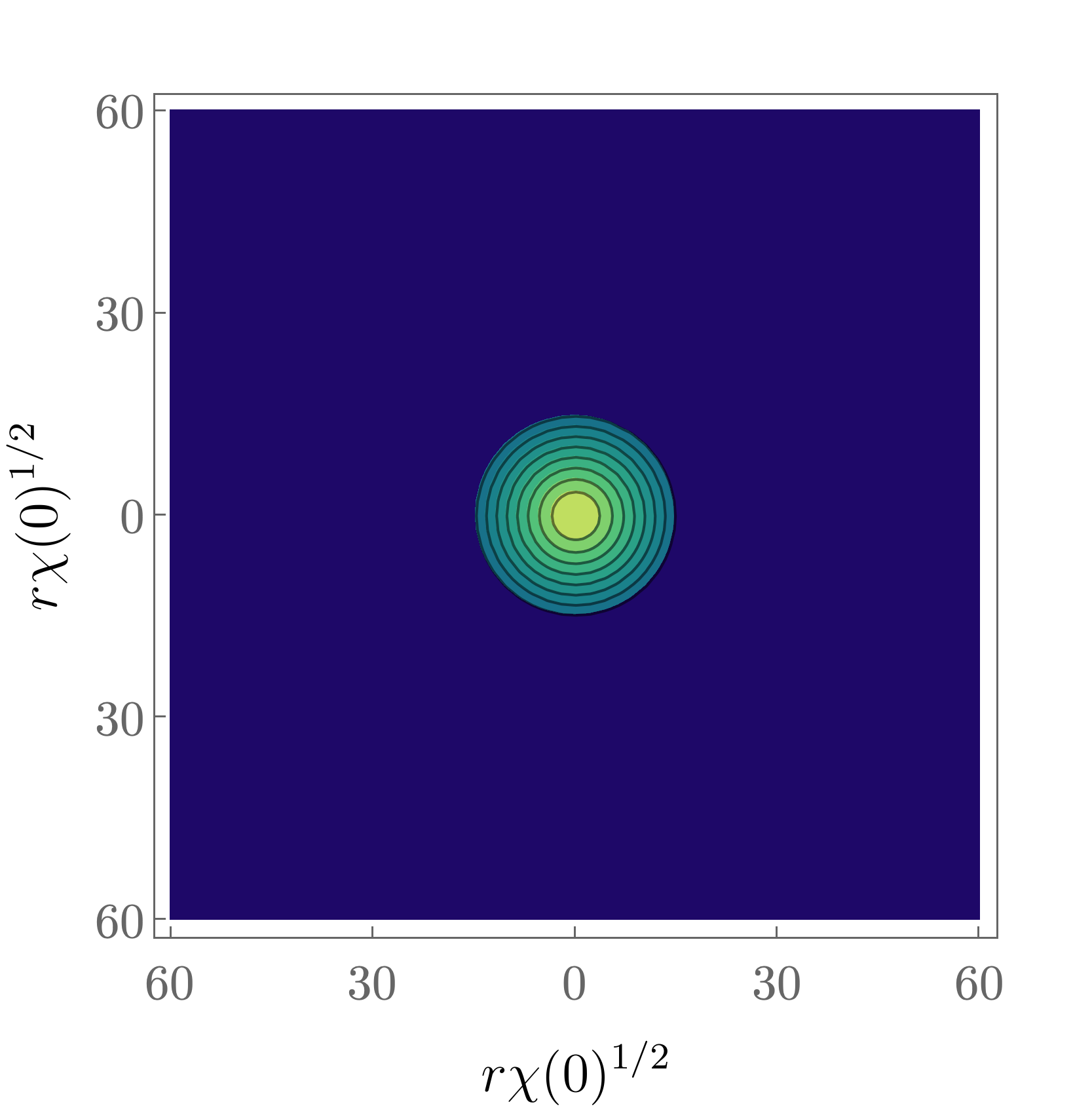}
        \includegraphics[width=0.47\textwidth]{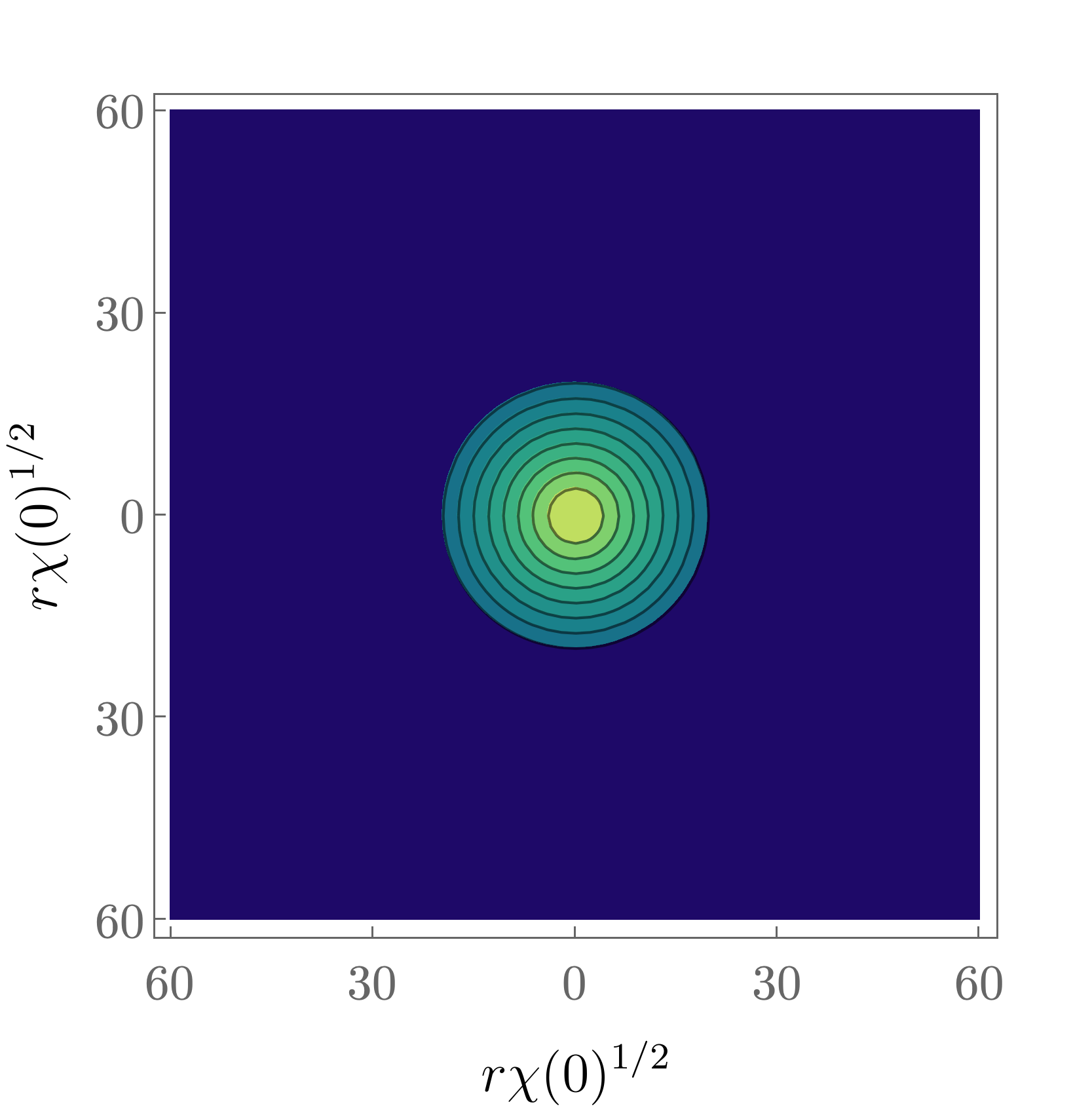}
        \includegraphics[width=0.47\textwidth]{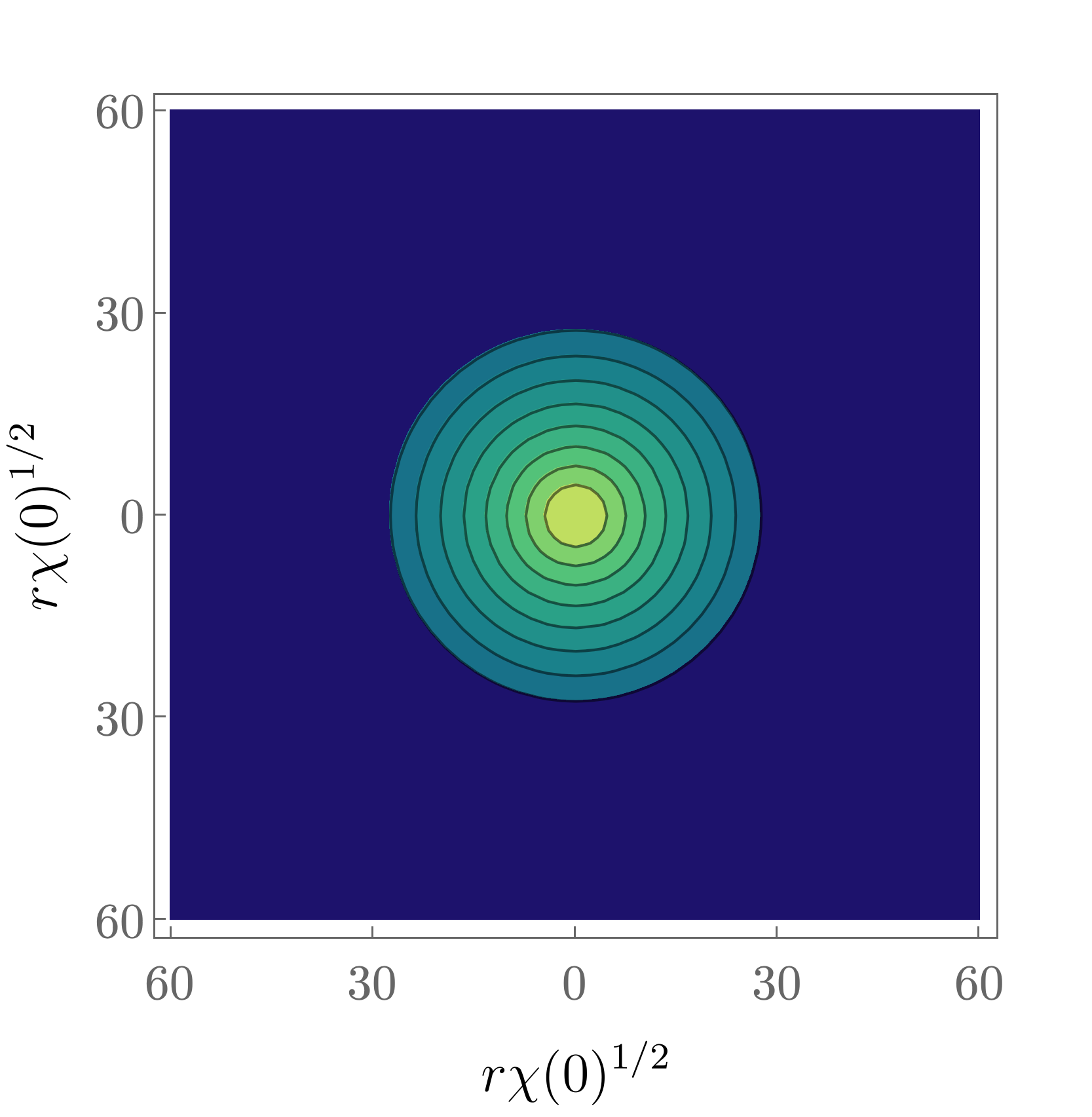}
        \includegraphics[width=0.47\textwidth]{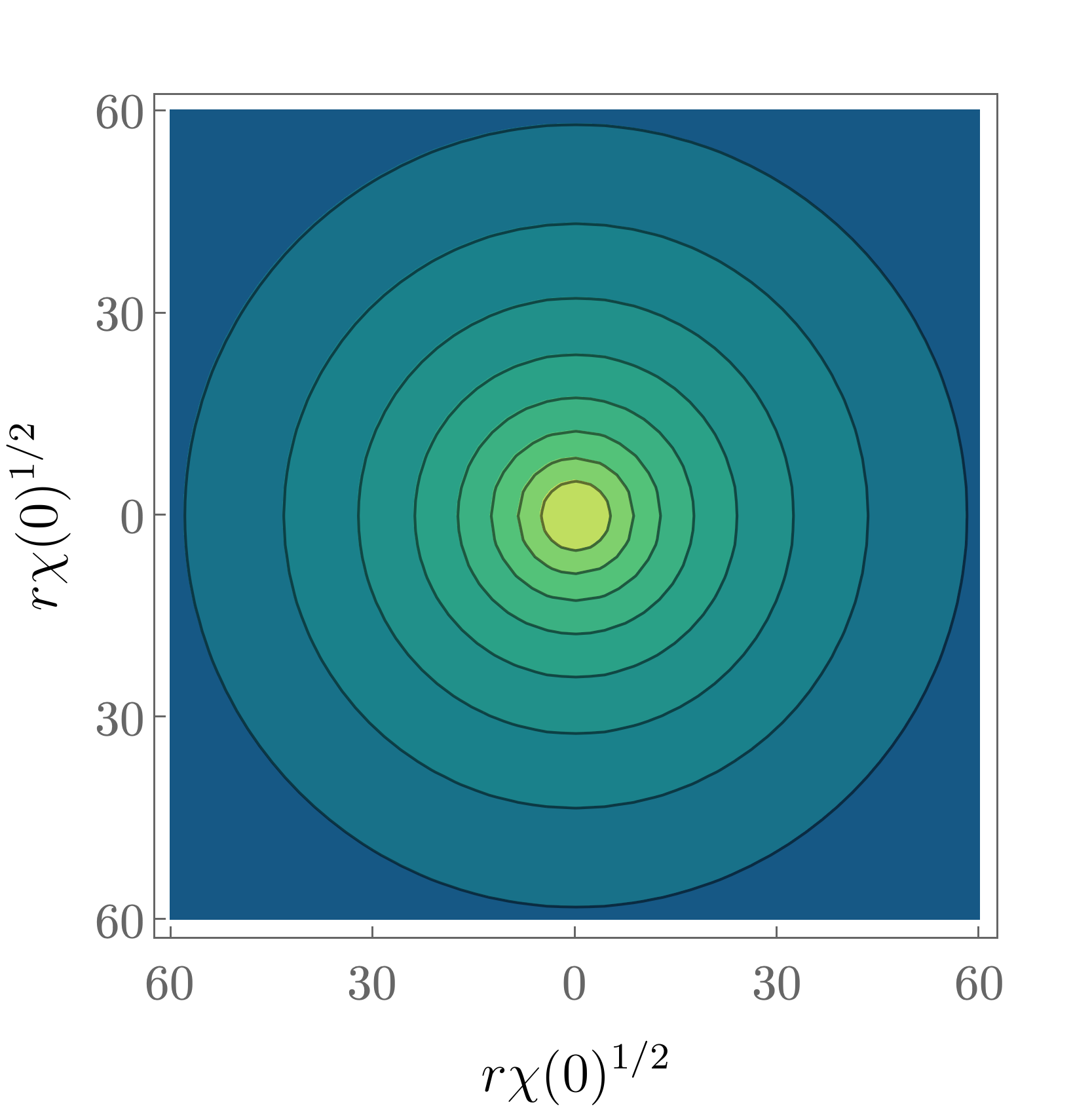}
    \end{center}
    \caption{\small Puffing-up of the HP ball in $d=6$. We show contour plots of $\log_{10}\chi$, for a fixed value $\chi(0)$, for $d = 3, 4, 5, 6$. The ball size grows as $d$ increases and  $\Delta_{\beta}$ decreases, i.e., the temperature approaches the Hagedorn limit from below. In $d=3,4,5$ the ball is well localized, with $\chi$ falling off exponentially. In $d=6$ the fall-off is instead a power-law. To more clearly visualize this effect, we plot $\log_{10}\chi$. In the four plots the contour lines are at the same values of $\chi$.}
      \label{fig:puff-up-balls}
\end{figure}

For $d\geq 6$ it can be proven that \eqref{eq:eigensystem} do not have normalizable ball solutions whenever $\Delta_\beta>0$ \cite{Horowitz:1997jc,Chen:2021dsw}. The physical reason is that the gravitational interaction in these dimensions is too short-ranged to hold together a ball condensate, and when attempting to do so, the ball puffs-up. This phenomenon, which will play a role later on, is consistent with the finding in \cite{Balthazar:2022hno} of ball solutions in $d=6$ with $\Delta_\beta=0$ and a power-law fall-off. In Fig.~\ref{fig:puff-up-balls} we illustrate how the HP balls gradually become less strongly localized as the dimension grows, 
asymptotically behaving like
\begin{equation}\label{eq:chiexp}
     \chi\sim r^{\frac{1-d}{2}} e^{-\sqrt{\Delta_\beta}\, r}\,,
\end{equation}
in $d=3,4,5$ and puffing up in $d=6$ since they only exist with $\Delta_\beta=0$. 
However, corrections to the effective theory invalidate the Hagedorn-temperature solutions and reintroduce balls with $\Delta_\beta>0$ \cite{Balthazar:2022hno}. Their significance is not fully clear yet and in this article we will not study them.

Eq.~\eqref{eq:chiexp} also shows that we can identify the radius of the ball as
\begin{equation}\label{rball}
    r_b=\frac1{\sqrt{\Delta_\beta}}=\sqrt{\frac{T}{T_H-T}}\,.
\end{equation}
Note that this radius increases with temperature, which contrasts sharply with the behavior of black holes, where the horizon radius $r_0=(d-2)/(4\pi T)$ decreases as the temperature rises. Later we will identify additional differences between HP balls and black holes that stem from this distinct behavior.

\paragraph{Thermodynamics.} Let us now see how we can obtain the relation $\beta(M)$ and then $S(M)$ for the string ball states. Observe from \eqref{asymppot} that the combination 
\begin{equation}
    \frac{G_N M}{\Delta_{\beta}^{(4-d)/2}}\equiv \textsf{g}_d^{-(4-d)/2}
\end{equation}
is a pure number (in string units) that is invariant under the rescaling \eqref{eq:scaling}. Therefore $\textsf{g}_d$ is an invariant in the mass-temperature relation for all the balls in a given $d$.
If we compute it in some (arbitrary) reference solution $(\varphi_0(r),\chi_0(r))$ with mass $M_0$ and temperature $\Delta_{\beta 0}$,
then in any other solution the mass and temperature are related by
\begin{equation}\label{GMDelta}
   G_N M =G_N M_0\left(\frac{\Delta_\beta}{\Delta_{\beta 0}}\right)^\frac{4-d}{2}=\left(\frac{\Delta_\beta}{\textsf{g}_d}\right)^\frac{4-d}{2}\,,
\end{equation}
or conversely,
\begin{equation}\label{eq:betaM}
    \frac{\beta}{\beta_H}=1+\textsf{g}_d(G_N M)^\frac{2}{4-d}\,.
\end{equation}
If we now integrate the first law of thermodynamics we obtain the entropy
\begin{equation}
    S(M)=\int \beta(M) dM\,,
\end{equation}
and in this manner we recover \eqref{SMHPb} (where we absorbed Newton's constant in $G_N^\frac{2}{4-d} \textsf{g}_d \to \textsf{g}_d$). The reason why we can compute entropy in a solution to classical field equations is that the mass of the condensate, $m^2_\infty=\Delta_\beta$,  measures the size of the Euclidean time circle and therefore is linked to a temperature. Although the same scaling symmetry is present for a boson star, in that case $m^2_\infty$ bears no relation to a temperature.

From our numerical analysis, we have obtained
\begin{equation}
    \textsf{g}_3 = 0.325539\,,\qquad \textsf{g}_5 = 1068.28\,.
\end{equation}
For $d=4$ the scaling argument breaks down since the value of $G_N M$ is itself invariant under rescalings, which is a consequence of the scale invariance of the Newtonian system in this dimension. Then, all the string balls have the same mass (in string units) regardless of their size and temperature, and we have found it to be
\begin{equation}\label{GMd4}
    G_N M = 12.0871\,.
\end{equation}
As a result, we cannot obtain the self-gravitational corrections to the free-string entropy within the approximations of the HP model. Nevertheless, string balls of different sizes have different temperatures: $\Delta_\beta$ scales linearly with the amplitude of the ball, and our numerical solution gives the proportionality factor,
\begin{equation}
    \Delta_\beta=0.199627\,\chi(0)\,.
\end{equation}
This scaling is the reason why we can meaningfully refer to these solutions as HP balls and not black holes, even though $G_N M >1$ suggests that the string ball could be within its Schwarzschild radius: We can always scale the ball up to a size larger than the black hole radius by raising the temperature closer to the Hagedorn limit.\footnote{If $\Delta_\beta\sim 1$ then the ball transitions to a black hole, and if $\Delta_\beta\lesssim g^2$ then it will become an almost free string ball. See \eqref{eq:qlimit} and the discussion below.}

The free energy of these balls can now be obtained using the expression \eqref{eq:freeF}, which gives
\begin{equation}
    F=\frac{2}{6-d}\textsf{g}_d^\frac{d-4}{2}\Delta_\beta^\frac{6-d}{2}\,.
\end{equation}
This result, which has good scaling behavior, is the same as \eqref{eq:Fball} near the Hagedorn limit and is appropriate in all $d<6$.

Local thermodynamic stability, diagnosed by the sign of the specific heat $C=dM/dT$, is well-known to signal that the Schwarzschild black hole, which has $C<0$, evaporates into radiation. For HP balls, the specific heat from \eqref{eq:betaM} is
\begin{equation}\label{eq:Cball}
    C=\frac{d-4}{2 \textsf{g}_d}\frac{T_H}{T^2}\left(\frac{\textsf{g}_d T}{T_H-T}\right)^{\frac{d-2}{2}}\,.
\end{equation}
Since HP balls in $d=3$ have $C<0$, they are naturally expected to decay by thermally radiating with a temperature close to $T_H$. In $d=4$ the infinite specific heat of the Hagedorn gas remains unresolved by gravitational interactions in this approximation, but the string is expected to emit radiation at any non-zero $g$. In $d=5$, HP balls have $C>0$, which may seem to suggest that they will remain stable. This conclusion is misleading. As we will discuss shortly, in this dimension the HP balls coexist with black holes and (almost) free-string balls of the same mass and larger entropy. The HP ball is then globally unstable to evolve into these states, which will decay by emitting radiation whenever the coupling $g$ is not zero.


Finally, observe that \eqref{eq:betaM} relates the mass and radius of an HP ball as
\begin{equation}\label{eq:Mrb}
    G_N M\sim r_b^{d-4} \qquad \textrm{(HP~balls)}\,,
\end{equation}
to be contrasted with the behavior of black holes with horizon radius $r_0$,
\begin{equation}\label{eq:Mr0}
    G_N M \sim r_0^{d-2} \qquad \textrm{(black~holes)}\,.
\end{equation}
We see that  in $d=3$ HP balls become more compact the heavier they are, and in this respect, they act like conventional stars and unlike black holes. In $d=5$ HP balls are more similar to black holes, whose size grows with the mass.

\paragraph{HP balls, black holes, and free strings.} Finally, to connect with the correspondence principle, we restore string units and write the entropy formula \eqref{SMHPb} in the form
\begin{equation}
    \frac{S}{\beta_H M}=1+\frac{d-4}{d-6}\textsf{g}_d\frac{\alpha'}{\kappa}\left(\frac{G_N M}{\alpha'/\kappa}\right)^\frac{2}{4-d}\,.
\end{equation}
For the sake of this discussion, we neglect any unnecessary purely numerical factors and write Newton's constant as $G_N \sim g^2 M_s^{1-d}$, where $g$ is the string coupling and $M_s =1/\sqrt{\alpha'}$ is the string mass. Then the self-interaction correction to the free string entropy is
\begin{equation}
    \frac{S}{\beta_H M}-1
    \propto \frac{d-4}{d-6}\left(g^2 \frac{M}{M_s}\right)^\frac{2}{4-d}\,.
\end{equation}
We see that the size of the corrections is measured by the value of
\begin{equation}
    g^2 \frac{M}{M_s}\approx g^2 S\,,
\end{equation}
which is the 't~Hooft-like coupling for fundamental string states introduced in  \cite{Ceplak:2023afb}. Black holes have $g^2 S \gg 1$, while self-gravitating strings have $g^2 S<1$. The transition is expected to happen when $g^2 S \sim 1$, but as we have seen, this is imprecise in $d = 4$ since all the HP solutions that interpolate between almost free strings and black holes have the same mass and entropy. In general, the transition is  characterized by $\Delta_\beta\sim 1$, that is, $r_b\sim \ell_s$.

When $d=3$ the relative increase in the free string entropy is a small amount $\sim (g^2 S)^2$. For $d=4$, we saw that there are no corrections to the order we are working on. For $d=5$ they are $\sim (g^2 S)^{-2}$, and moreover, negative, so the HP ball has lower entropy than a free string ball.

As we explained, whenever the string coupling is non-zero, quantum fluctuations set an upper bound  \eqref{eq:qlimit} on the temperature for which HP solutions are valid. For HP balls this means a lower bound on their mass,
\begin{equation}\label{eq:Mbound}
    \frac{M}{M_s}\gtrsim g^{-\frac{4}{6-d}}\,,
\end{equation}
or in the 't~Hooft coupling,
\begin{equation}\label{eq:g2Sbound}
    g^2 S \gtrsim S^{\frac{d-4}{2}}\,,\quad \textrm{i.e.,}\quad g^2 S \gtrsim g^{2\frac{4-d}{6-d}}\,.
\end{equation}
The classical limit of the HP model is $g\to 0$, $S\to\infty$, with $g^2 S$ finite. At any non-zero value of $g$, if \eqref{eq:g2Sbound} is violated the classical description of the ball breaks down because the configuration puffs up into a free string ball. In $d=3$ this happens when $g^2 S\sim g^{2/3} \ll 1$ and the free string ball is far from the regime $g^2 S\gtrsim 1$ where black holes exist. In $d=4$, the free string with $g^2 S\sim O(1)$ has much bigger size than a black hole, but these can exist with masses not much larger. In $d=5$ the free string phase appears in a regime of $g^2 S\sim 1/g^2 \gg 1$ where the HP balls and free strings coexist with more entropic black holes \cite{Chen:2021dsw}.

\subsection{HP strings in a circle}
\label{sec:HP-strings}

Now we turn to study HP solutions in a space $\mathbb{R}^{d-1} \times S^1$ where one of the directions is compactified into a circle of length $L$,
\begin{equation}
    z\sim z+L\,.
\end{equation}
The scaling symmetry now acts like
\begin{equation}\label{eq:scalingL}
    (x^i,\chi,\varphi,\Delta_\beta, L)\to(\lambda^{-1/2} x^i,\lambda\chi,\lambda\varphi, \lambda\Delta_\beta, \lambda^{-1/2} L)\,,
\end{equation}
and therefore relates solutions with different amplitudes, sizes and temperatures which live in circles of different length. We will use
\begin{equation}
    \textsf{L} =L\sqrt{\Delta_\beta}=\frac{L}{r_b}
\end{equation}
as a scale-invariant measure of the circle length. Doing this, the unit of length is the thickness $r_b$ of the string \eqref{rball} (which is also the Compton wavelength of the thermal scalar \eqref{eq:m2}).

Maintaining spherical symmetry in the directions transverse to $z$, the equations to solve are
\begin{equation}
    \begin{aligned}
        \partial_r^2 \chi + \frac{d-2}{r}\partial_r \chi+ \partial_z^2 \chi-\left(\Delta_\beta-\varphi\right)\chi & =0,\\
        \partial_r^2\varphi + \frac{d-2}{r}\partial_r\varphi + \partial_z^2 \varphi - \frac12 \chi^2 & = 0,
    \end{aligned}
    \label{eq:non-uniform}
\end{equation}
with $\chi$ and $\varphi$ periodic in $z\in [-L/2,L/2]$.
We can immediately construct uniform configurations by translating the HP balls along $z$, and we will refer to them as UHP strings, or simply UHPS. Their entropy \eqref{SMHPs} is easily obtained.

We are more interested in non-uniform solutions.
An indication of their existence comes from the finding in \cite{Chen:2021dsw} that localized string balls have negative Euclidean modes, i.e., they admit off-shell deformations that decrease the value of the Euclidean action.\footnote{More precisely, \cite{Chen:2021dsw} found deformations that decrease the action, but they did not isolate the negative eigenmode. We do this below.} A general argument, well-known for black holes \cite{Gregory:1987nb,Reall:2001ag}, connects these negative Euclidean modes of the localized solution to the presence of a zero mode of the uniform string.\footnote{Namely, an on-shell perturbation $\propto e^{ik z}$ of the uniform solution in $d$ dimensions is the same as an off-shell eigenmode of the fluctuation operator for the localized solution in $d-1$ dimensions, with negative eigenvalue $-k^2$. See \eqref{eq:linpert} below.}
This zero mode is a static deformation that creates non-uniformity along the length of the string. It is a small, linearized perturbation, which signals the appearance of a family of non-uniform string configurations, namely the non-linear extensions of the zero mode.

Our goal is to numerically construct and study these branches of non-uniform HP strings, which we refer to as NUHPS. 
But before we delve into their fully non-linear construction, we will find the zero mode.

\subsubsection{Zero mode}\label{subsubsection:zeromode}

We take the solution $(\chi_s(r),\varphi_s(r))$ for a localized ball in $d-1$ dimensions that we have numerically obtained in Sec.~\ref{sec:HP-balls}, and uniformly extend it along the direction $z$. We then perturb it as 
\begin{equation}
    \chi(r,z) = \chi_s(r)+\delta \chi_0(r)e^{-i k z}, \qquad \varphi(r,z) = \varphi_s(r)+\delta \varphi_0(r)e^{-i k z}\,.
\end{equation}
Plugging this into \eqref{eq:non-uniform} and linearizing in the perturbation we get
\begin{equation}\label{eq:linpert}
\begin{aligned}
     \delta \chi_0'' + \frac{d-2}{r} \delta \chi_0' - \left(\Delta_{\beta}+ \varphi_s\right)\delta \chi_0  - \chi_s \delta\varphi_0= &  k^2 \delta \chi_0, \\
    \delta \varphi_0'' + \frac{d-2}{r} \delta \varphi_0'   -  \chi_s \delta \chi_0= & k^2\delta \varphi_0\,,
\end{aligned}
\end{equation}
which is an eigenvalue problem for $(\delta \chi_0,\delta\phi_0)$ with eigenvalue $k^2$. Here $\Delta_{\beta}$ is the value for the solution $(\chi_s,\varphi_s)$. 

This problem can be easily solved numerically to find $k$. The scale-invariant combination
\begin{equation}\label{Lstar2}
    \textsf{L}_*=\frac{2\pi}{k} \sqrt{\Delta_\beta}=\frac{2\pi}{k r_b}
\end{equation}
measures the length of the circle that supports the zero mode creating non-uniformity on a string of thickness $r_b$.

For the same reason that localized HP balls only exist in $d=3,4,5$, uniform HP strings are possible only when $d=4,5,6$. For these cases, our numerical results give values of $\textsf{L}_*$ that decrease with $d$, namely,
\begin{equation}\label{L0HP}
    \textsf{L}_*= 7.95516,\ 4.8416,\ 2.76554\,,
\end{equation}
which is also the generic behavior with $d$ seen in black strings \cite{Gregory:1993vy,Emparan:2018bmi}. Let us compare this to the wavelength of the zero mode for black strings. There, it is convenient and common to use the horizon radius $r_0$ to define a scale-invariant critical length,
\begin{equation}
    \textsf{L}_*= \frac{2\pi}{k r_0}\,.
\end{equation}
In the same three dimensions as above, black strings have
\begin{equation}\label{L0BS}
    \textsf{L}_*= 7.17127,\ 4.95162,\ 3.9748\,.
\end{equation}
The comparison between \eqref{L0HP} and \eqref{L0BS} is appropriate since in both cases we are measuring $L$ relative to the thickness of the string, $r_b$ or $r_0$.\footnote{The correspondence between these results is sensible when not only the black hole size but also $L$ are around the string scale. Otherwise, when $L\gg 1$ the two zero modes happen for very different masses, see Fig.~\ref{fig:all_phases}.} The similarity of the results is notable considering that they arise from quite different equations.\footnote{The HP ball in $d=6$ would give a uniform HP string in $d=7$ which also possesses a zero mode. The correct solution must include corrections the HP effective theory , but if we still do not include them, then, normalizing with the amplitude of $\chi$, the results in $d=4,5,6,7$ are $L_*\sqrt{\chi(0)}=9.56146$,\,$10.8363$,\,$12.3143$,\,$14.0784$, i.e., they increase with $d$. Further work including corrections to the theory would be desirable.}

The presence of the zero mode suggests that UHP strings should exhibit a GL instability, and it is natural to expect that this happens when they become longer than fatter, i.e., when  $\textsf{L} > \textsf{L}_*$. While the dynamically unstable modes cannot be found within the HP formalism, the study of the properties of non-uniform static solutions will give us evidence for them.

\subsubsection{Non-uniform string solutions}

Now we turn to finding general string configurations with non-uniformity beyond the linearized approximation. This requires the numerical solution of an eigenvalue problem for \eqref{eq:non-uniform}, for a configuration with arbitrarily chosen amplitude. 

This time, we must solve the problem for a range of values of $L$, so as to obtain the eigenvalue $\Delta_\beta$ as a function of $L$. Once we find the solutions, the total mass is also a function of $L$ given by 
\begin{equation}\label{eq:MNUHPS}
    M(L)=-\frac{(d-3)\omega_{d-2}}{8\pi G_N} \lim_{r\to\infty} r^{d-3}\int_{-L/2}^{L/2} dz\, \varphi(r,z;L)\,.
\end{equation}

Besides compactifying the radial direction $r\to u$ as in \eqref{eq:rtou}, it is convenient to work with a redefined coordinate $x = 2z/L$ which ranges over $[-1,1]$. Then $L$ enters as a parameter in the equations. If we require symmetry $x\rightarrow -x$ then we can restrict to $x \in [0,1]$ and the reflected solution will be periodic in the $x$ circle.

In these coordinates, the asymptotic boundary conditions are 
\begin{equation}
    \chi(1,x)=\varphi(1,x)=0\,, 
\end{equation}
while regularity at the fixed-points of the reflection requires
\begin{equation}
    \partial_x \chi\vert_{x=0,1} = \partial_x \varphi\vert_{x=0,1}=0\,.
\end{equation}
Regularity along the rotational symmetry axis, $\partial_r \chi\vert_{r=0}=0$, is automatically implemented when using the $u$ coordinate.
Finally, we select the reference solutions by setting $\chi(0,0) = 1$, again an arbitrary but simple and harmless choice.
With these conditions, we have an eigenvalue problem for $\Delta_\beta$ for each value of $L$. Once again, we discretize the domain of $(u,x)$ with a Lobatto-Chebyshev grid and use a relaxation method to find the family of non-uniform solutions.

We start by using the zero mode as a seed for the relaxation method, to generate solutions with length $L$ slightly larger than $L_*$ and small non-uniformity. In this way, we can construct a family of solutions with increasing $L$ and growing non-uniformity, until the zero mode is no longer a good seed. Then we continue by using the last solution as a seed for the next one at a slightly bigger $L$. In this manner, we progressively generate NUHPS for larger $L > L_*$ and increasing non-uniformity. 

\begin{figure}
    \begin{center}
        \includegraphics[width=0.47\textwidth]{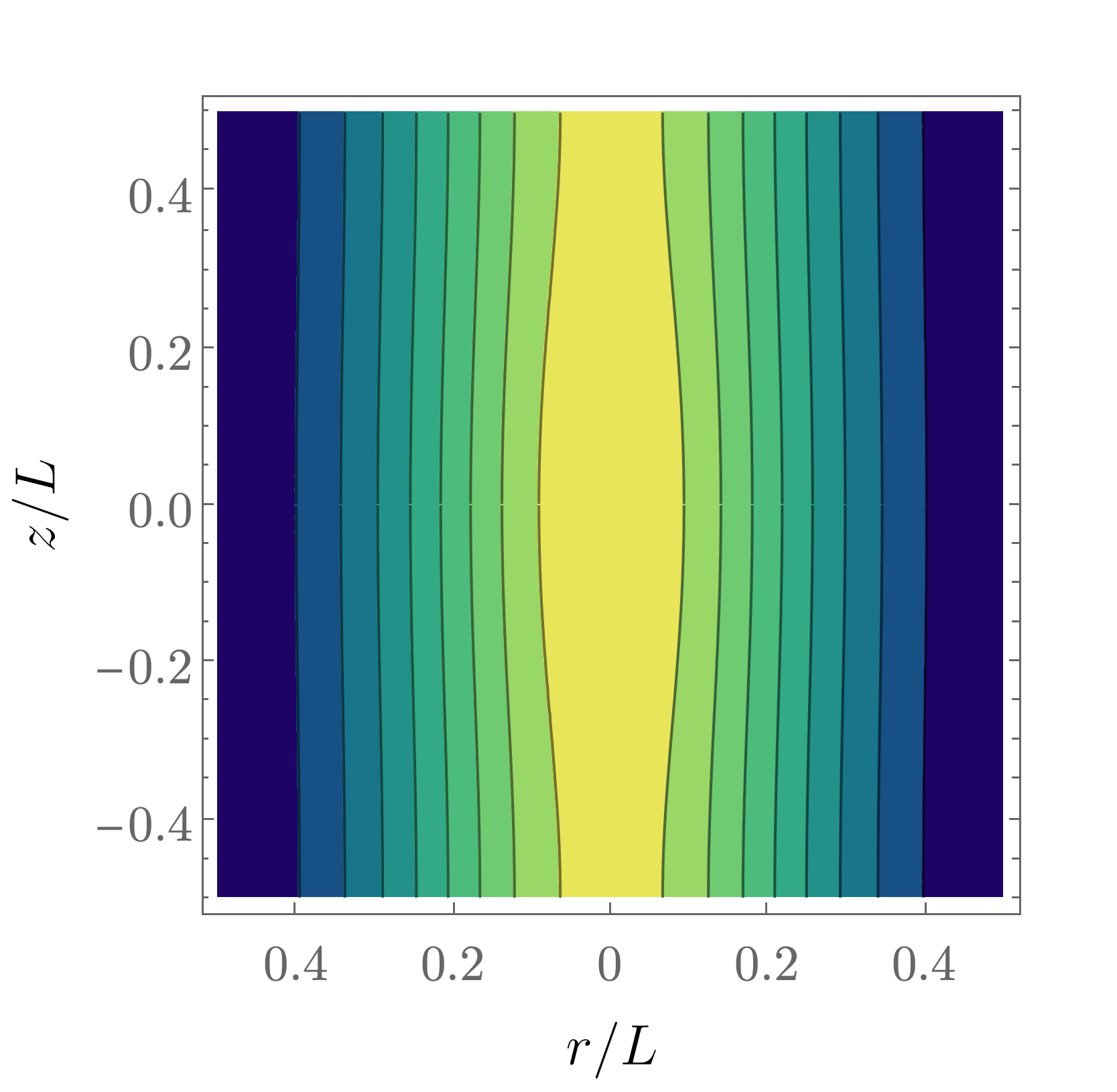}
        \includegraphics[width=0.47\textwidth]{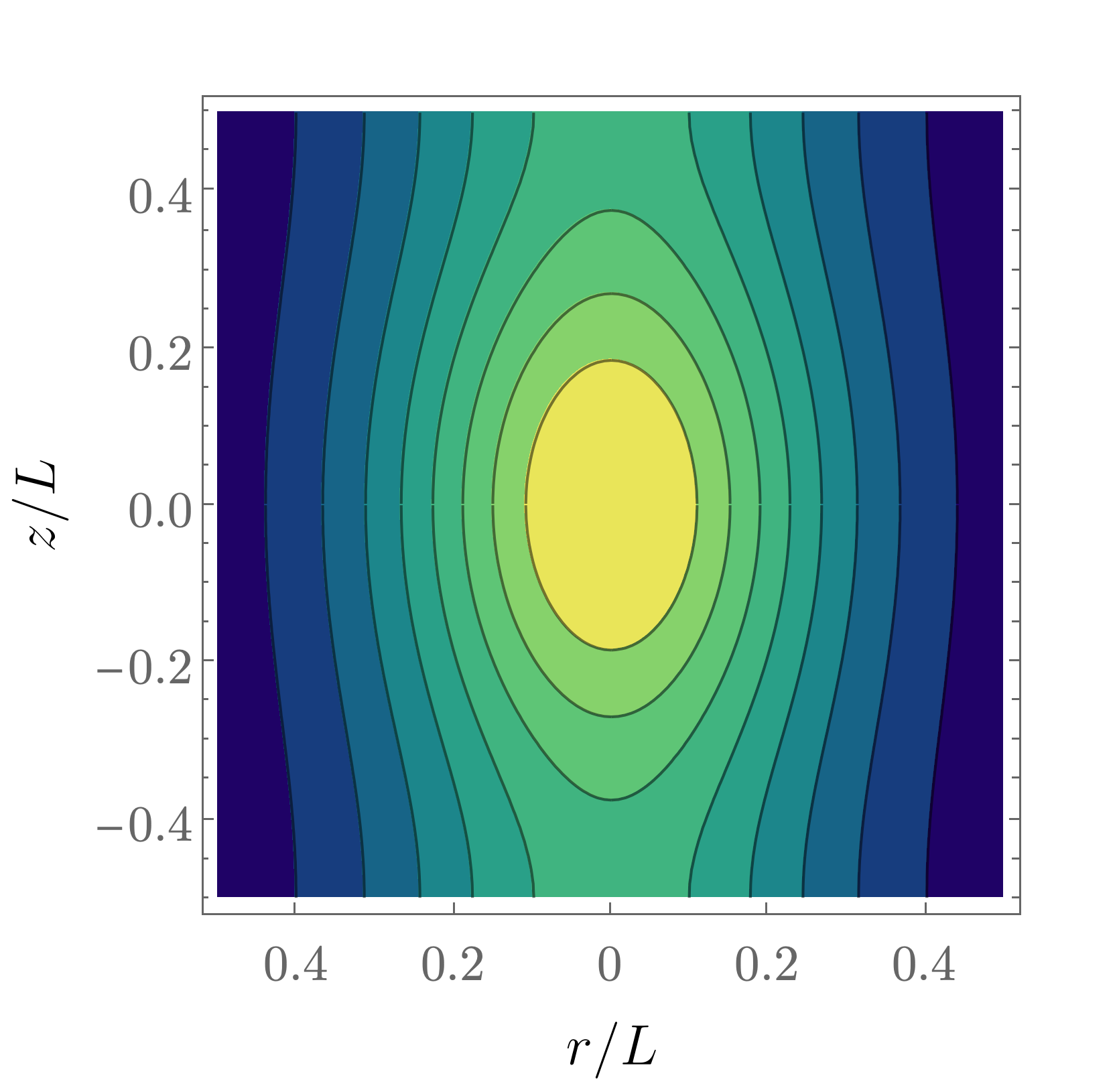}
        \includegraphics[width=0.47\textwidth]{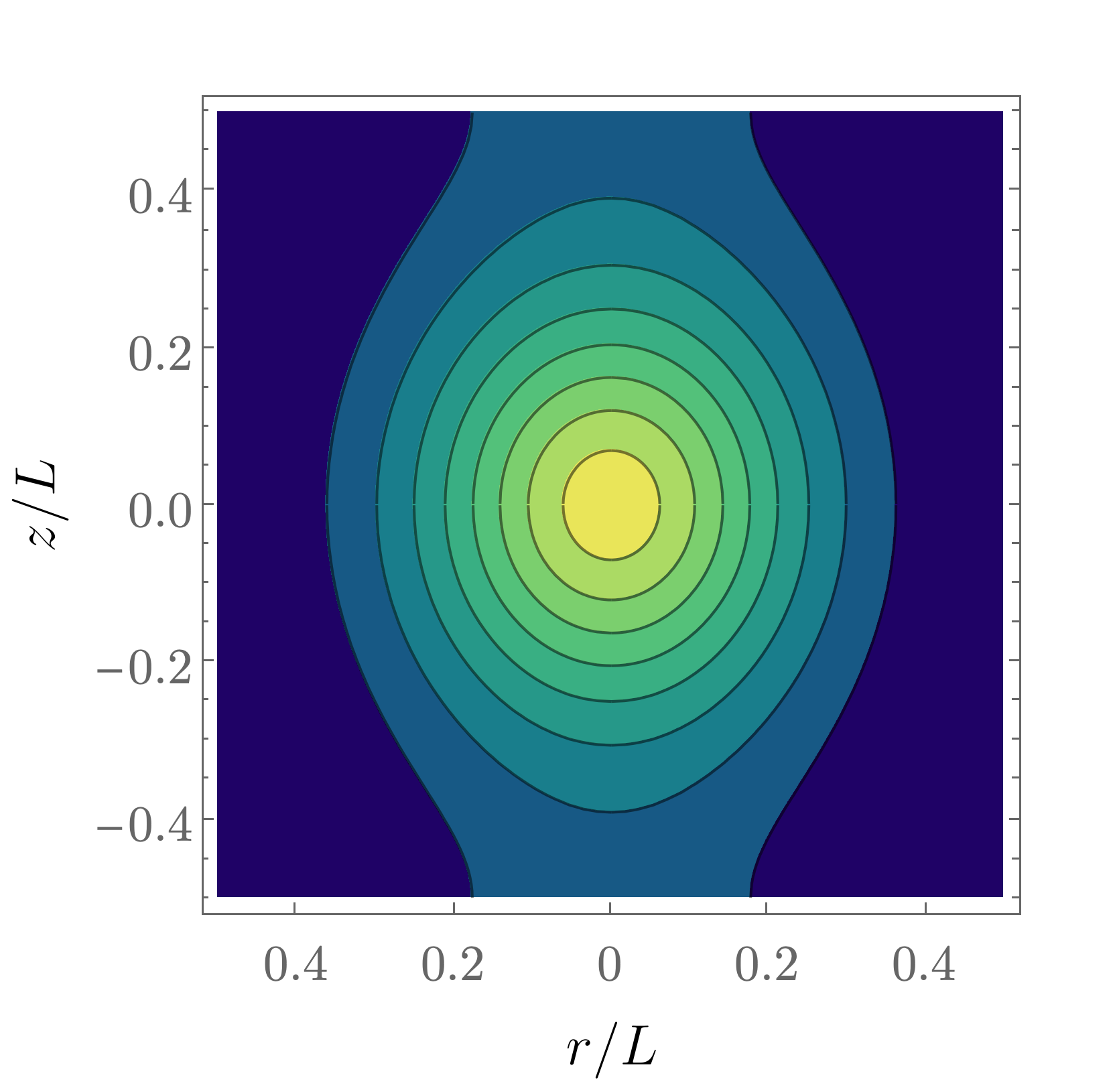}
         \includegraphics[width=0.47\textwidth]{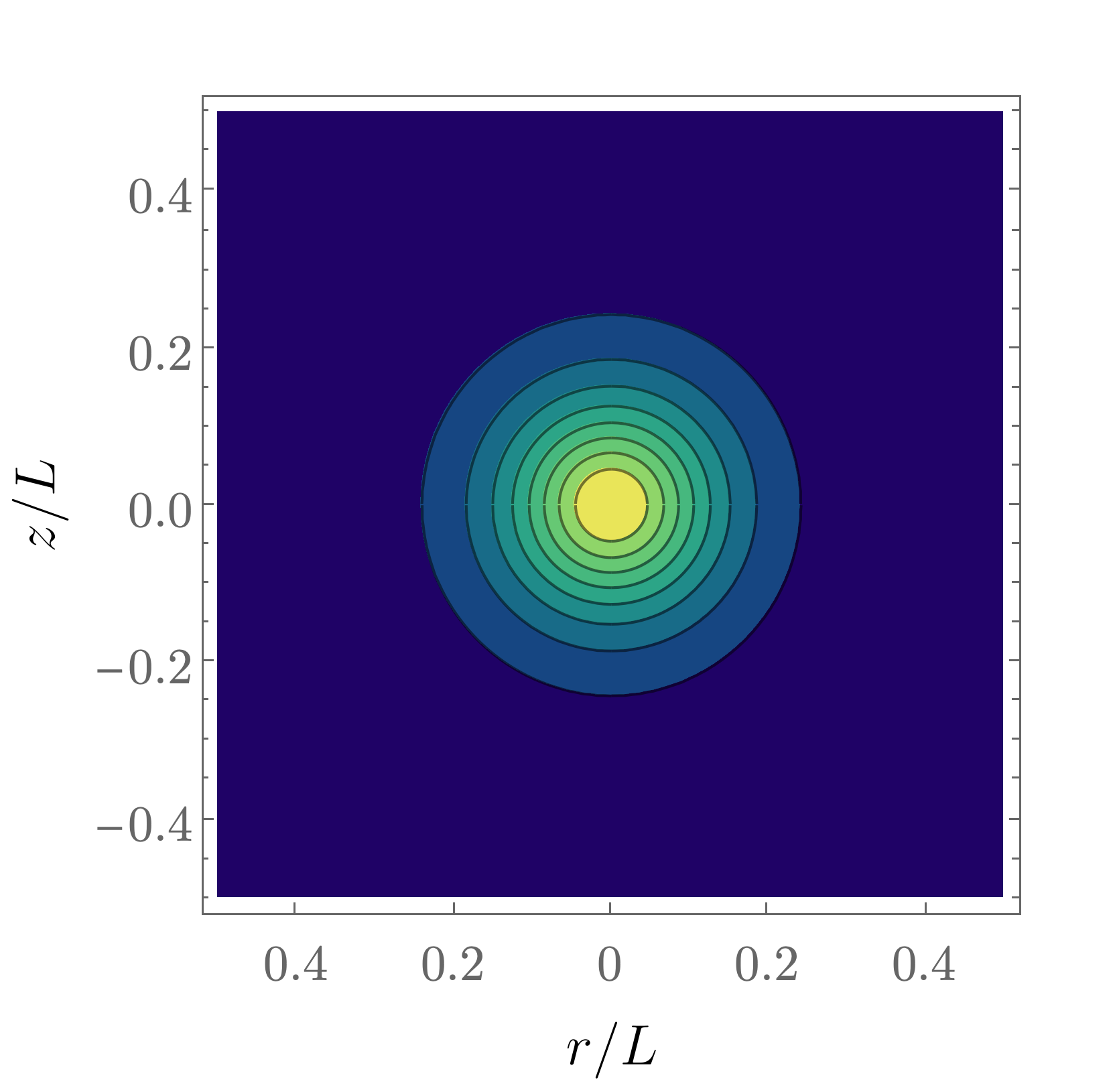}
    \end{center}
    \caption{\small Contour plots of $\chi$ for the non-uniform string configurations in $d=5$ for increasing $\textsf{L} =L\sqrt{\Delta_\beta} = $
    4.842, 4.845, 4.909, 6.978 (the zero mode wavelength is $\textsf{L}_*=4.8416$). As the invariant length grows, the string quickly localizes and approaches a localized ball. For fixed $L$ we can view this as decreasing the temperature of the condensate, which reduces its size $r_b$ and in this dimension also means decreasing its mass (see Fig.~\ref{fig:M_vs_L}).}
      \label{fig:HP-perturbative-3D}
\end{figure}

We have verified that when $L$ becomes very large, keeping the amplitude of the condensate $\chi(0,0)$ fixed, the solutions we find approximate the localized, spherical string balls we found in Sec.~\ref{sec:HP-balls} (see Fig.~\ref{fig:HP-perturbative-3D}). This is a limit where  $\textsf{L}\gg 1$, so we could also imagine that we are keeping $L$ fixed and then decrease the temperature so that $\Delta_\beta$ grows. As we do this, the ball shrinks, but when its temperature is much less than the Hagedorn temperature, the effective HP theory is not valid anymore. The proper way to recover the ball is keeping the condensate size and $\Delta_\beta$ finite (and small), and then letting $L\to\infty$.

This connection between the HP string and the HP ball is only possible in the dimensions where both exist, that is, in $d=4,5$. In $d=3$ there are no uniform HP strings to begin with, and although in this dimension there exist localized balls when $L\to \infty$, the potential for any finite $L$ would grow logarithmically at $r\gg L$.

\begin{figure}
    \begin{center}
        \includegraphics[width=0.47\textwidth]{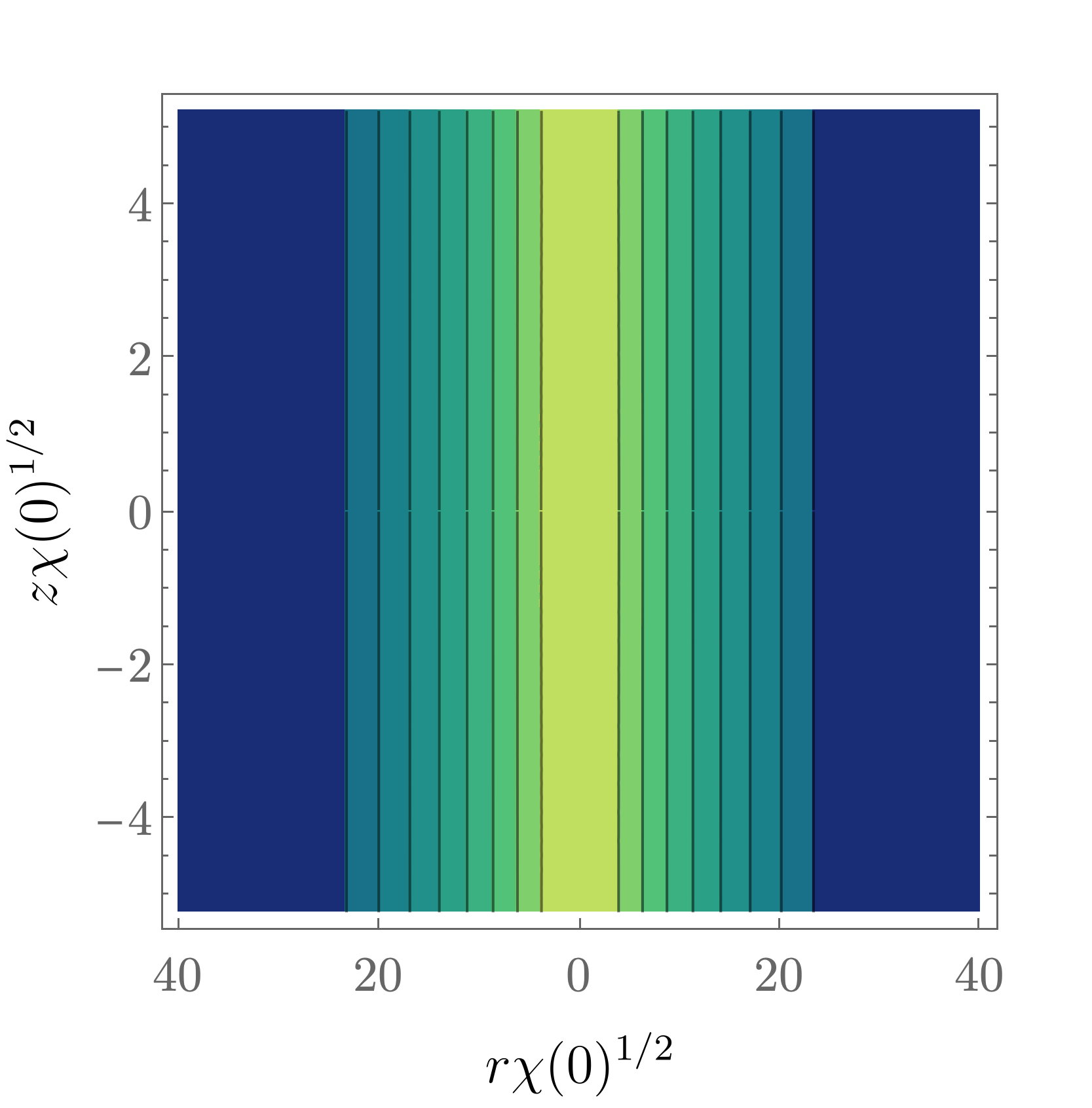}
        \includegraphics[width=0.47\textwidth]{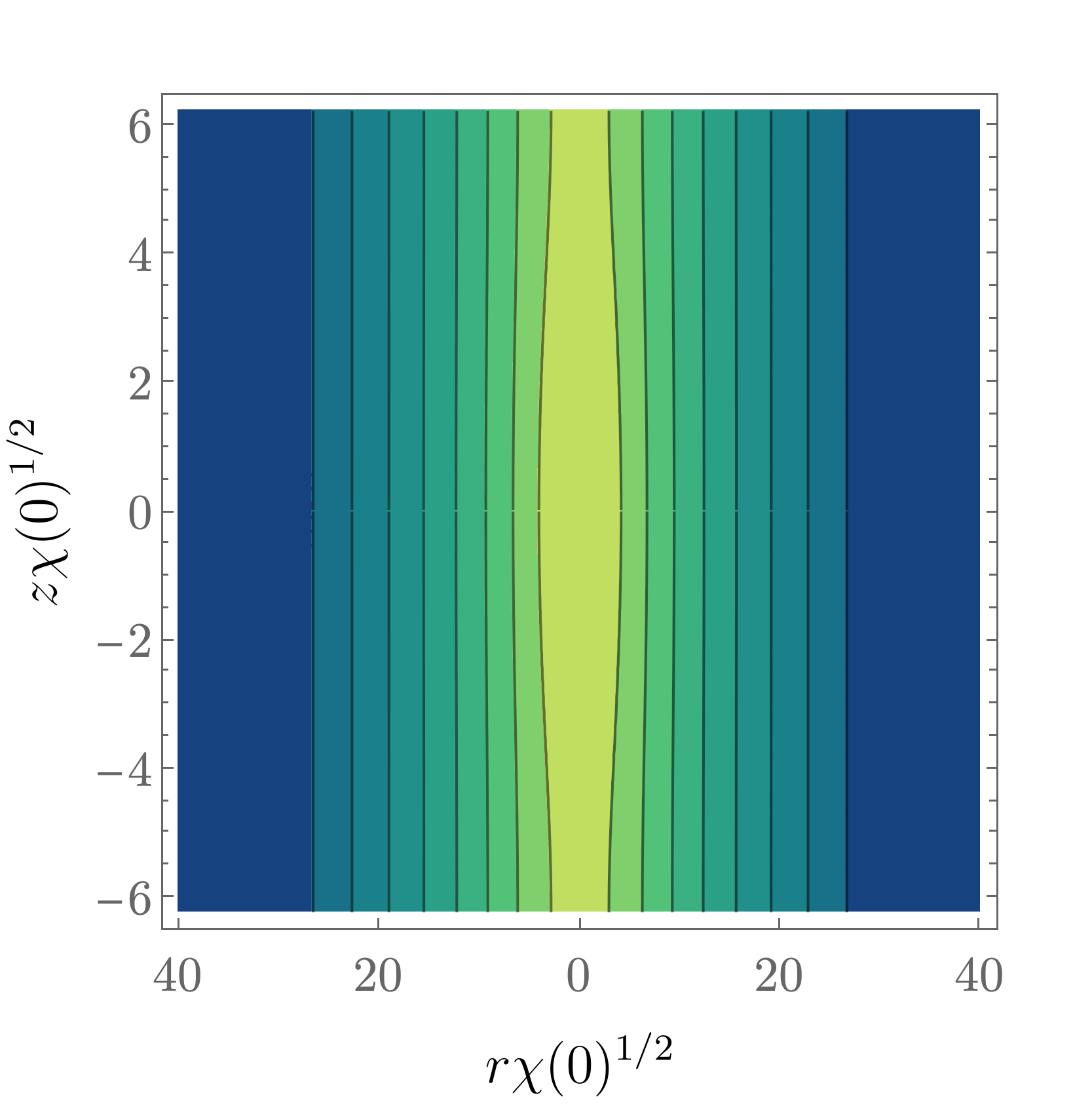}
        \includegraphics[width=0.47\textwidth]{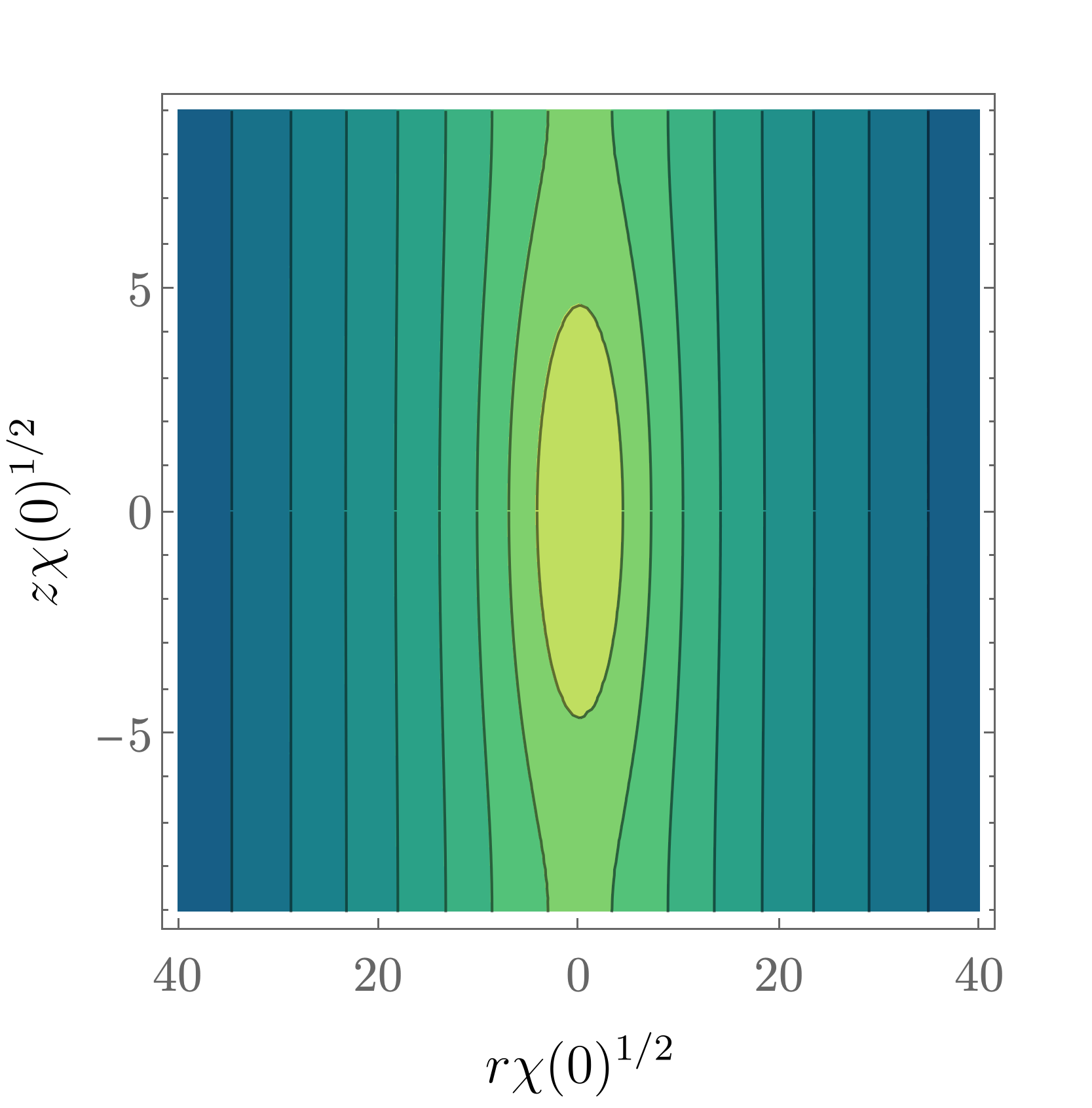}
         \includegraphics[width=0.47\textwidth]{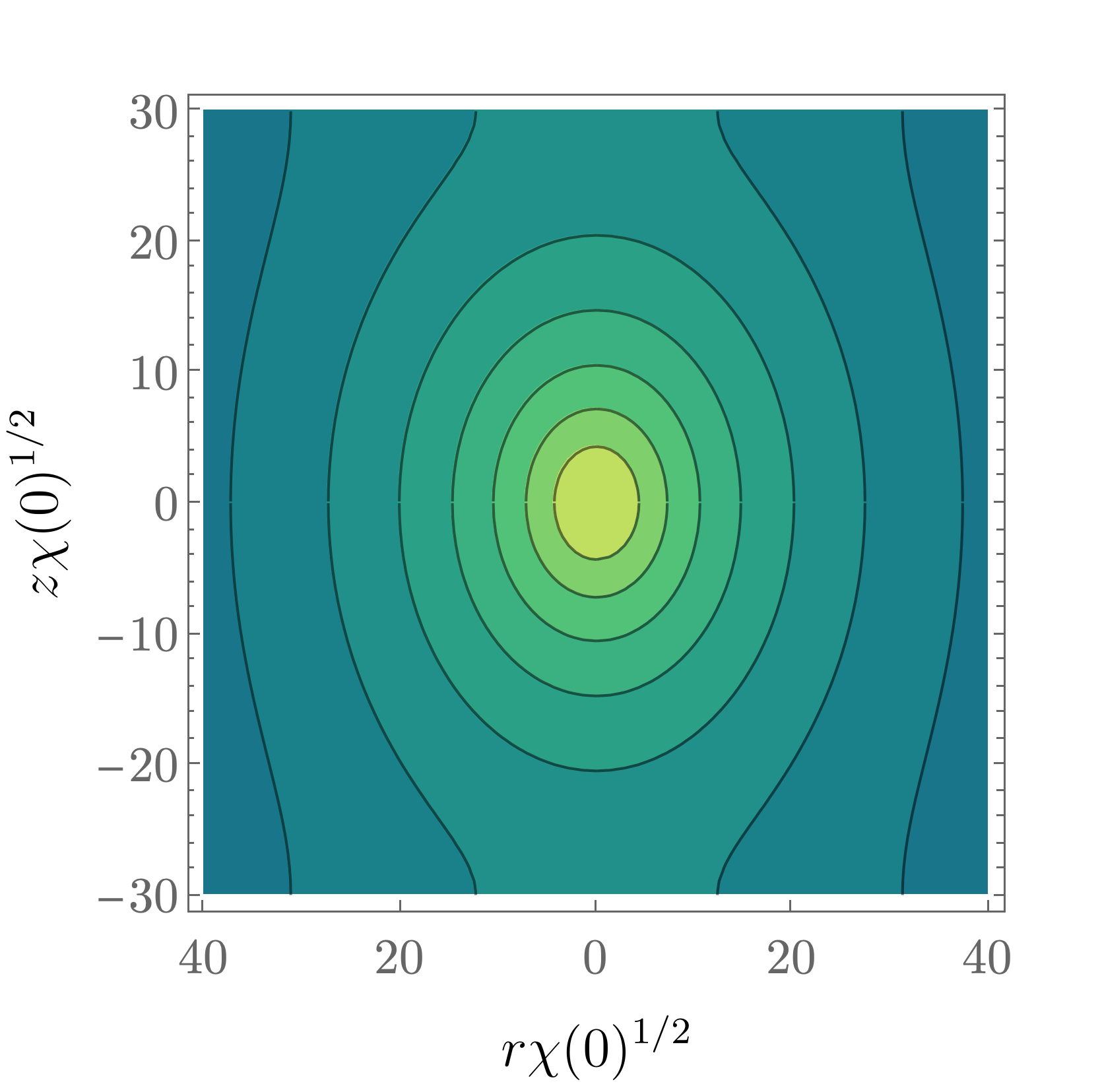}
    \end{center}
    \caption{\small Puffing-up of non-uniform strings in $d=6$ as the non-uniformity grows. We show contour plots of $\log_{10}\chi$, for a fixed value $\chi(0,0)$, for decreasing $\textsf{L} =L\sqrt{\Delta_\beta} = 2.76536, 2.7066, 2.34838, 0.94902$ 
    (the zero mode length is $\textsf{L}_*=2.76554$). 
    As we reduce the scale-invariant length, the string quickly localizes in the compact circle direction $z$, while it expands in the transverse radial direction $r$. Although $\textsf{L}$ decreases as the inhomogeneity grows, $L$ increases (observe the range in the vertical axis) while  $\Delta_{\beta}$ decreases faster, i.e., the temperature quickly approaches the Hagedorn limit from below and the size $r_b$ grows very large. We plot $\log_{10}\chi$ to visualize more clearly the power-law (non-exponential) radial decay of  $\chi$.}
      \label{fig:puff-up-d-6}
\end{figure}

In $d=6$ the situation is reversed. As $L$ is made larger and the non-uniformity of the HP configuration grows, we find that  $\Delta_\beta$ approaches zero and the condensate spreads farther out. In the limit $L\to\infty$, $\chi$ does not fall off exponentially 
but as a power law (see Fig.~\ref{fig:puff-up-d-6}). 
We have verified that the configuration approaches the ball solution of \cite{Balthazar:2022szl}, not via their `dimensional regularization' but in a sequence of physical states. The subsequent study in \cite{Balthazar:2022hno} suggests that when the non-uniformity of our solutions grows sufficiently large, the corrections to the HP effective theory will become important\footnote{We thank Jinwei Chu for emphasizing this point.}. While it should be interesting to study these effects, for the time being we will continue discussing the solutions within the realm of the leading-order HP theory. We observe that as the non-uniformity increases, the mass grows large, and we can say that the string condensate puffs up. As we explained, this is a consequence of the shorter range of the gravitational interaction in $d=6$. It is curious to note that in this case, although $L$ increases with growing inhomogeneity, the scale-invariant length $\textsf{L}$ decreases since $\Delta_{\beta}$ decreases faster than $L$ grows, i.e., the temperature approaches very quickly the Hagedorn limit from below.

\subsection{Scaling and entropy}

We can now apply the scaling symmetry to derive the entropy $S(M,L)$, now adapted to take into account that the length of the circle must be scaled too.

Say that we have obtained a reference solution $\varphi_0(r,z;L_0)$, $\chi_0(r,z;L_0)$ in a circle of length $L_0$, with mass $M_0$ and temperature $\Delta_{\beta 0}$. Then, from \eqref{eq:MNUHPS} we see that the scaling symmetry relates it to other solutions through
\begin{equation}
    \frac{M(L)}{M_0(L_0)}=\left(\frac{\Delta_{\beta}}{\Delta_{\beta 0}}\right)^\frac{4-d}{2}\,,\qquad \frac{L}{L_0}=\sqrt{\frac{\Delta_{\beta 0}}{\Delta_{\beta}}}\,.
\end{equation}
We deduce that the mass $M$ of a solution in a circle of length $L$ depends on the temperature $\beta$ as
\begin{equation}
    M(\beta,L)=\left(\frac{\Delta_{\beta}}{\Delta_{\beta 0}}\right)^\frac{4-d}{2}M_0\left(L\sqrt{\frac{\Delta_{\beta}}{\Delta_{\beta 0}}} \right)\,.
\end{equation}
Since $M_0$ depends on $\beta$ through its dependence on the circle length, we cannot in general invert this analyticaly to solve for $\beta(M,L)$. However, this can easily be done numerically, and then the entropy follows by integrating the Clausius relation
\begin{equation}\label{eq:clausius}
    \frac{\partial S(M,L)}{\partial M}=\beta(M,L)\,.
\end{equation}

To provide an integration limit for $M$, we take the mass of the uniform string $M_*(L)$ with length $L$ equal to the wavelength of the zero-mode. This is adequate, since the uniform and non-uniform branches meet at that solution, and the properties of uniform solutions are easily obtained from those of balls as we will explain now.

We can find the mass $M_*(L)$ by scaling the mass $M_{0*}(L_{0*})$ of the critical solution in the reference family. Since
\begin{equation}\label{L*}
    \textsf{L}_*=L_{0*} \sqrt{\Delta_{\beta0*}}=L\sqrt{\Delta_\beta}\,,
\end{equation}
we get
\begin{equation}
    M_*(L)=\left(\frac{\Delta_{\beta}}{\Delta_{\beta 0*}}\right)^\frac{4-d}{2}M_{0*}\left(\frac{\textsf{L}_*}{\sqrt{\Delta_{\beta 0*}}}\right)\,.
\end{equation}
The necessary data from the reference solution are given by $\textsf{g}_{d-1}$ and $\textsf{L}_*$. Eq.~\eqref{GMDelta} allows us to write the mass of a uniform string as
\begin{equation}
    G_NM=L\left(\frac{\Delta_\beta}{\textsf{g}_{d-1}}\right)^\frac{5-d}{2}\,,
\end{equation}
which at the critical length, \eqref{L*}, becomes
\begin{equation}\label{MstarL}
    G_N M_*(L) =\left(\frac{\textsf{L}_*^2}{\textsf{g}_{d-1}}\right)^\frac{5-d}{2} L^{d-4}\,.
\end{equation}

This mass and length correspond to the configuration at the bifurcation point, so the entropy of the non-uniform strings must match that of the uniform strings,
\begin{equation}
    S(M_*(L),L)=S_U(M_*(L))\,.
\end{equation}
Since the uniform string is obtained by translating a $(d-1)$-dimensional string ball along the length $L$, its entropy is
\begin{equation}\label{eq:SUM}
    S_U(M_*(L))=\beta_H M_*(L) + \beta_H M_*(L)\,\textsf{g}_{d-1} \frac{d-5}{d-7} \left(\frac{G_N M_*(L)}{L}\right)^{\frac{2}{5-d}}\,
\end{equation}
 (cf.~\eqref{SMHPs} in Sec.~\ref{sec:basics}), which, we note, can be written as
 \begin{equation}
     S_U(M_*(L))=\beta_H M_*(L)\left(1+\frac{d-5}{d-7}\left(\frac{\textsf{L}_*}{L}\right)^2\right)\,.
 \end{equation}
Now we can integrate \eqref{eq:clausius},
\begin{equation}
    S(M,L)=S_U(M_*(L))+\int_{M_*(L)}^M dM'\,\beta(M',L)\,.
\end{equation}
Using \eqref{eq:SUM} and $\Delta_\beta$ instead of $\beta$, we can also write it as
 \begin{equation}\label{eq:entropy_NUS}
     S(M,L)=\beta_H M + \beta_H  M_*(L)\, \textsf{g}_{d-1} \frac{d-5}{d-7} \left(\frac{G_N M_*(L)}{L}\right)^{\frac{2}{5-d}}+\beta_H\int_{M_*(L)}^M dM'\,\Delta_\beta(M',L)\,.
 \end{equation}
Observe that since $\Delta_\beta\geq 0$, the self-interaction corrections to the free string entropy in $d=4,5$ are always positive. However, in $d=6$ the sign of the corrections cannot be inferred so easily.

The function $S(M,L)$ can be readily obtained by numerical integration of the reference solutions. 
Then, using \eqref{eq:freeF} we can also compute the free energy $F(\beta,L)$.

\section{Thermodynamics of HP phases}\label{sec:entropy_results}

The results for the different phases in $d = 4,5,6$ and their thermodynamics are shown in Figs.~\ref{fig:M_vs_L}, \ref{fig:M_vs_L_L_normalized}, \ref{fig:s_vs_m_non_uniform} and \ref{fig:f_vs_minfty_non_uniform}. In the plots, the non-uniform strings are shown as solid black curves and uniform strings in green (solid and dashed---more below). In $d=6$ we also show, red dot-dashed, the entropy $S=\beta_H M$ of free strings so puffed-up that their self-gravity is negligible, since it is relevant for determining the microcanonical dominant phase. Free string states do not play a role in the canonical ensemble for $\beta>\beta_H$, since, within the approximations we work in, they are always at the Hagedorn temperature. Other than this, the results we show are for HP balls and strings in the classical regime of the HP theory, with $g\to 0$ and finite $g^2 S$, but it should be understood that when $g$ is small but non-zero, the solutions below the range in \eqref{eq:qlimit} must be replaced by almost-free strings. In $d=6$, the solutions with very large non-uniformity are expected to receive significant corrections from higher-order interactions in the effective theory.

For the uniform strings we distinguish between solid-green and dashed-green according to whether $\textsf{L}=L\sqrt{\Delta_\beta}=L/r_b$ is smaller or bigger than the critical value $\textsf{L}_*$. The former are fat strings which are expected to be GL-stable, while the latter are thin strings expected to be GL-unstable. In Sec.~\ref{sec:stabcrit} we will present arguments for this interpretation.

\subsection{HP phases on a KK circle} 

\begin{figure}[t]
\begin{center}
    \includegraphics[width=0.435\textwidth]{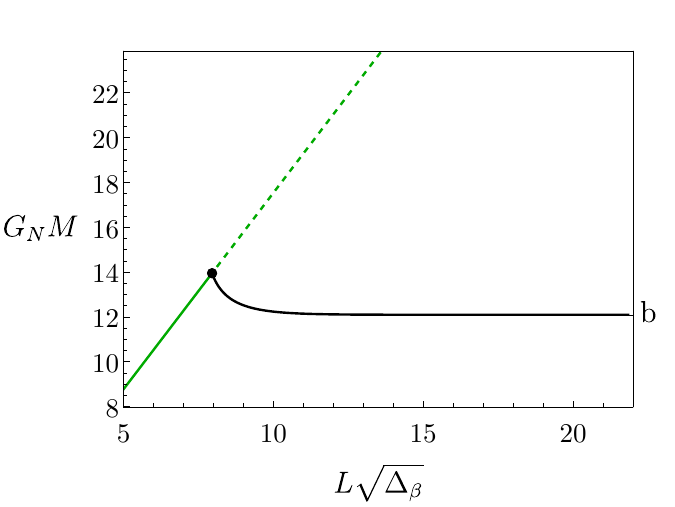}
    \quad 
    \includegraphics[width=0.5\textwidth]{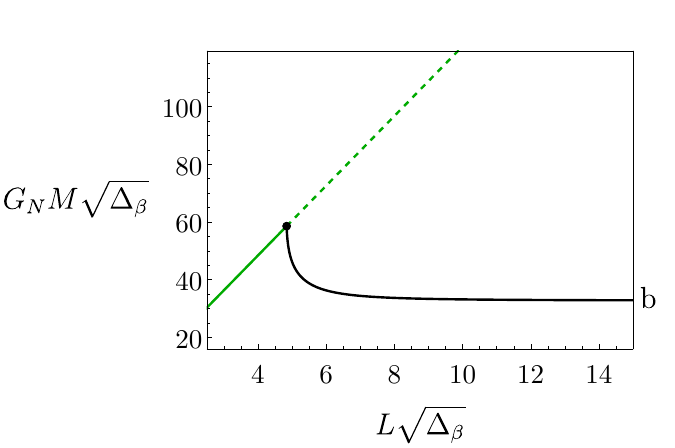}
    \includegraphics[width=0.47\textwidth]{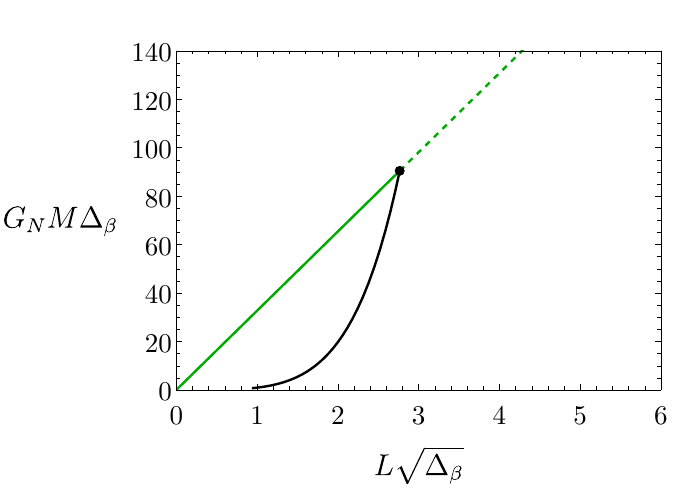}
\end{center}
\caption{\small 
Mass as a function of $L$, normalized to be scale-invariant for fixed $\beta$, in $d=4$ (top-left), $d=5$ (top-right), and $d=6$ (bottom). The non-uniform strings (solid black) branch off the uniform ones (green) when $\textsf{L}=L\sqrt{\Delta_\beta}$ reaches the zero mode value $\textsf{L}_*$ \eqref{L0HP}. Uniform strings are expected to be GL-unstable when thin enough, $L\sqrt{\Delta_\beta}>\textsf{L}_*$ (dashed green). On the rightmost vertical axis, ``b" denotes the mass of the localized string ball that NUHPS approach in $d=4,5$. 
}
  \label{fig:M_vs_L}
\end{figure}

\begin{figure}[t]
\begin{center}
    \includegraphics[width=0.475\textwidth]{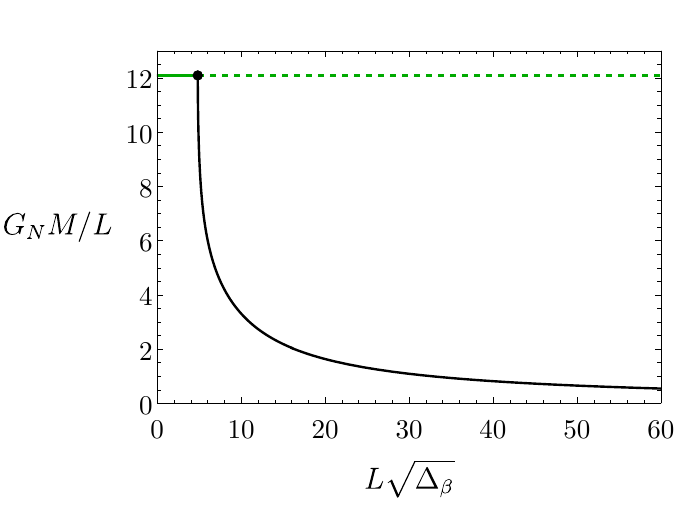}
    \quad
    \includegraphics[width=0.49\textwidth]{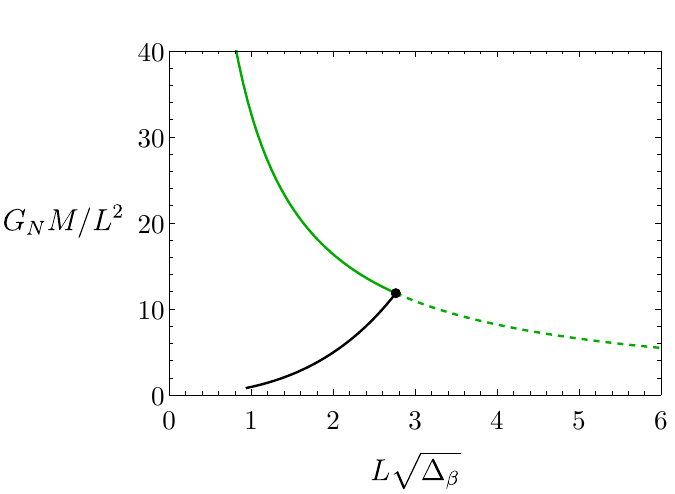}
\end{center}
\caption{\small Mass as a function of $\Delta_\beta=(\beta-\beta_H)/\beta_H$, normalized to be scale-invariant for fixed $L$, in $d=5$ (left) and $d=6$ (right). In $d=4$, since $G_N M$ is scale-invariant, the top-left panel in Fig.~\ref{fig:M_vs_L} shows this information.  
}
  \label{fig:M_vs_L_L_normalized}
\end{figure}

In Figs.~\ref{fig:M_vs_L} and \ref{fig:M_vs_L_L_normalized} we show the mass of the solutions as a function of $L$ for fixed $\beta$, and conversely, as a function of $\beta$ for fixed $L$. 

We find that, for a given temperature and length, if there exist non-uniform solutions they always have lower mass than the uniform strings. However, their ranges of existence differ depending on the number of dimensions.

\paragraph{$\bullet$ $M(L)$ (Fig.~\ref{fig:M_vs_L}):} 
In $d=4,5$, for small $L$ only uniform strings exist and their mass grows linearly with $L$. When $L$ reaches the critical zero-mode wavelength \eqref{L0HP}, non-uniform strings branch off the uniform solutions. Their mass decreases with larger $L$ and quickly approaches the asymptotic mass of a localized ball. 

In $d=6$ the behavior is reversed. NUHPS exist from $L=0$ up to the bifurcation point, and for larger $L$ only uniform strings remain. It may be surprising that the solutions become more non-uniform as the circle length becomes smaller, but this is an effect of fixing the temperature. This behavior should be compared to what is observed in Fig.~\ref{fig:puff-up-d-6}. There, for fixed amplitude of the condensate the non-uniformity grows together with $L$, but $L\sqrt{\Delta_\beta}$ decreases since the temperature quickly approaches the Hagedorn limit (at which the effective theory breaks down). 

When $L$ becomes smaller than the string length, $L<1$, the effective theory breaks down. Therefore, in $d=6$ at any fixed temperature, the NUHPS phases should not be extended below $L\sqrt{\Delta_\beta}<\sqrt{\Delta_\beta}$, and never reach zero. In addition to this, as we mentioned above at any fixed $\beta$ the solutions with large enough non-uniformity must be corrected.

\paragraph{$\bullet$ $M(\beta)$  (Fig.~\ref{fig:M_vs_L_L_normalized}):}
In the figure we only show $d=5,6$. In $d=4$ the mass $G_N M$ is scale-invariant and there is no difference between keeping $L$ or $\beta$ fixed, so the plot in Fig.~\ref{fig:M_vs_L} serves here too.

If we fix $L$, then in $d=4,5$ the NUHPS exist only below a critical temperature (i.e., above a critical $\Delta_\beta$), and then with a mass less than the uniform phase. Reaching the localized ball would require lowering the temperature too much below the Hagedorn value, which is outside the validity of the effective theory (the ball size would be less than the string scale and the string would not have a near-Hagedorn density of states). 

In $d=6$ the NUHPS exist only above a critical temperature, closer to the Hagedorn limit than the uniform strings, and with lower mass. Since we are fixing $L$, as we keep approaching the Hagedorn limit the HP solutions puff up as shown in Fig.~\ref{fig:puff-up-d-6} and actually they should be modified by higher-order terms.

From these plots we can read the sign of the specific heat $C=dM/dT$.\footnote{The specific heat of the UHP string is the same as that of an HP ball in one less dimension, see \eqref{eq:Cball}.} We find that
\begin{alignat}{2}
    d&=4: \quad C_{UHP} &<0\,,    \quad C_{NUHP}&>0\,\\
    d&=5: \quad C_{UHP} & =\infty\,,  \quad C_{NUHP}&>0\,\\
    d&=6: \quad C_{UHP} &>0\,,    \quad C_{NUHP}&<0\,
\end{alignat}
Observe that the local thermodynamic stability of the uniform and non-uniform phases is reversed in $d=4$ and $d=6$. We will return to this point in Sec.~\ref{sec:stabcrit}.

\begin{figure}
\begin{center}
    \includegraphics[width=0.45\textwidth]{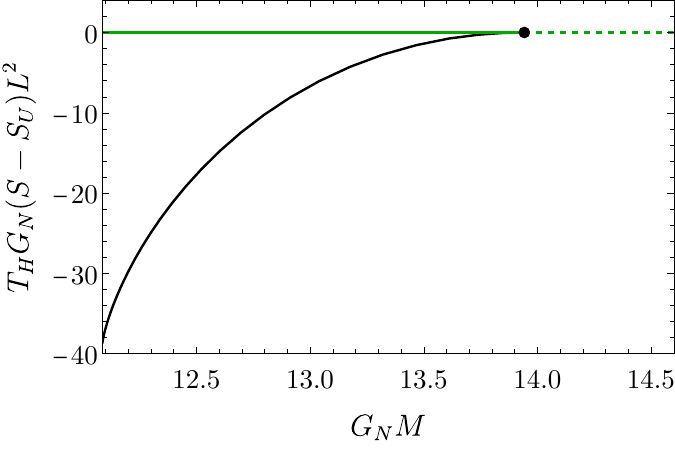}
    \qquad
    \includegraphics[width=0.45\textwidth]{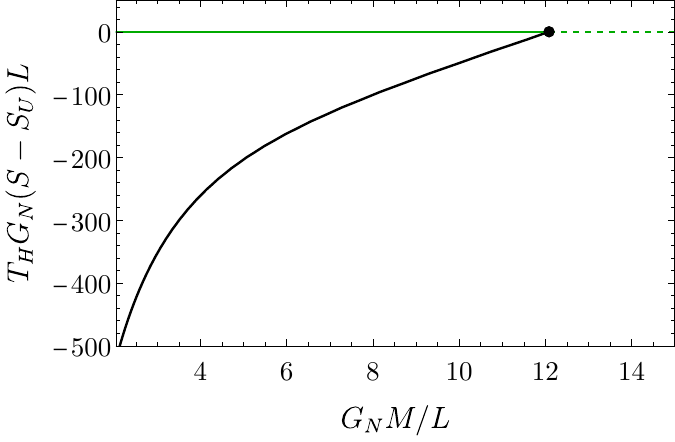}
    \includegraphics[width=0.45\textwidth]{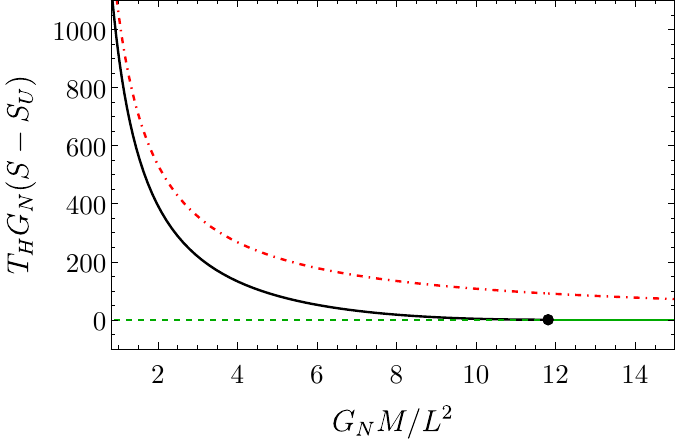}
\end{center}
\caption{\small Entropy of non-uniform HP strings (solid black) and free strings (red dot-dash) relative to uniform HP strings (green, zero line) at fixed length. Top-left: $d=4$. Top-right: $d=5$. Bottom: $d=6$. The axes correspond to scale-invariant magnitudes. The NUHPS bifurcate off at the critical $M_*(L)$ (black dot) towards lower masses. In $d=4,5$ the uniform phase dominates (at least until the mass is very small and they become free strings) but the unstable uniform strings (dashed green) likely evolve to a localized black hole (see Fig.~\ref{fig:all_phases}). In $d=6$, the non-uniform phase is preferred over the uniform one, but puffed-up free strings are always the overall dominant string phase..
}
  \label{fig:s_vs_m_non_uniform}
\end{figure}

\subsection[Microcanonical ensemble]{Microcanonical ensemble: $S(M,L)$ (Fig.~\ref{fig:s_vs_m_non_uniform})} 
We obtain the entropy of NUHPS by numerical evaluation of the integral in \eqref{eq:entropy_NUS}. To study the microcanonical ensemble, we fix the circle length $L$ and then vary the mass of the solutions. At a given mass, the configuration with the highest entropy should be the dominant phase. In $d=6$ we also consider a phase of almost-free strings, since in this dimension the string ball can puff up and become too weakly self-gravitating. In $d=4,5$ the range of the gravitational force is larger and, as we have seen, self-gravity actually gives a positive correction to the entropy so a free-string phase is not relevant unless the mass is very small, as we will discuss shortly.

In Fig.~\ref{fig:s_vs_m_non_uniform} we take the entropy of the uniform phase as a reference. We see that NUHPS bifurcate off the uniform solutions at a large enough value of the mass, and then the uniform and non-uniform phases coexist for lower masses.\footnote{For the UHPS in $d=5$ the value of $G_N M/L$ is equal to \eqref{GMd4}, i.e., their mass for fixed $L$ is a constant. In Fig.~\ref{fig:s_vs_m_non_uniform} we are naturally assuming that higher-order corrections to the effective theory yield a wider mass range for the UHPS. The effects in the entropy will be much smaller than the value of $S-S_U$ for the NUHPS.} In $d=4,5$ the NUHPS have lower entropy, and thus are subdominant in the microcanonical ensemble. As the mass decreases, they approach localized balls, but when they reach the lower mass limit in \eqref{eq:Mbound} they become gravitationally unbound string balls. For masses exceeding the critical zero-mode value, the UHPS seem to be the only viable phase. However, this overlooks the potential presence of black hole phases within this mass range. We will discuss this in Sec.~\ref{subsec:allphases}.

By contrast, in $d=6$ NUHPS have higher entropy than uniform ones below the critical zero-mode mass and might seem to be the dominant microcanonical phase. 
However, it is always entropically favorable for them to puff-up into a larger, almost-free-string phase. This is consistent with their negative specific heat, which signals a thermodynamic instability possibly indicating the transition into a free string phase. The properties of NUHPS at small mass and large non-uniformity are sensitive to higher-order corrections and must still be studied, but given the available evidence it seems plausible that they are also subdominant to free strings.

We conclude that the non-uniform HP phases are never thermodynamically dominant configurations in a KK circle for any given mass. 
This is consistent with the heuristic comparison in Sec.~\ref{sec:basics}, which concluded that localized HP balls have less entropy than uniform HP strings. 

\begin{figure}
\begin{center}
    \includegraphics[width=0.45\textwidth]{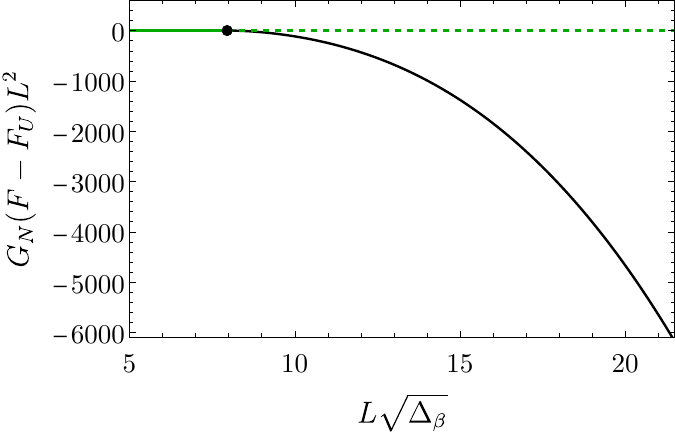}
    \qquad
    \includegraphics[width=0.45\textwidth]{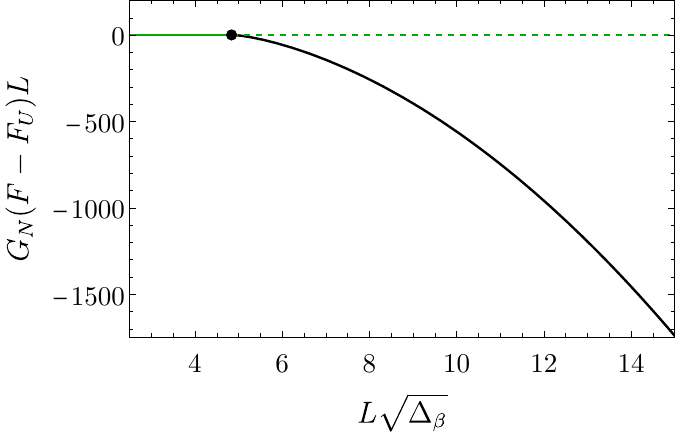}
    \includegraphics[width=0.45\textwidth]{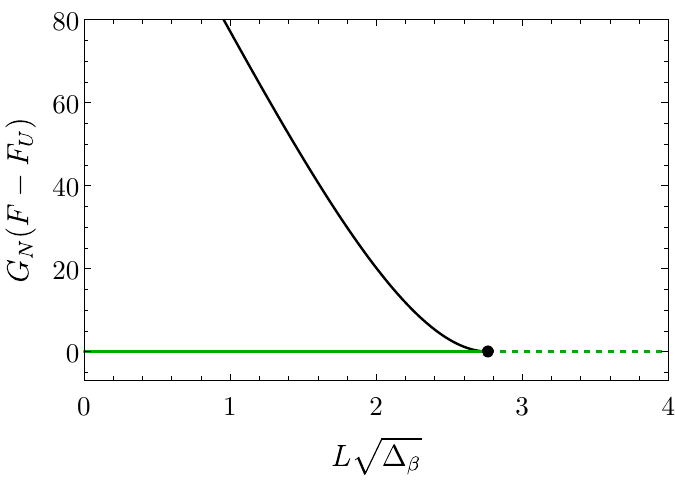}
\end{center}
\caption{\small Free energy density of non-uniform HP strings (solid black) and free strings (red dot-dash) at fixed circle length, as a function of the temperature, with $\Delta_\beta=(T_H-T)/T_H$ growing below the Hagedorn limit. Top-left: $d=4$. Top-right: $d=5$. Bottom: $d=6$. The axes correspond to scale-invariant magnitudes. The NUHPS bifurcate off at the critical $\textsf{L}_{*}$ (black dot) towards lower temperatures for $d=4$ and $5$, and towards the Hagedorn upper limit in $d=6$. In $d=4, 5$, the non-uniform phase dominates over the uniform one, and vice-versa in $d=6$. Free strings only play a role close enough to $T_H$, when the lower bound \eqref{eq:qlimit} on $\Delta_\beta$ is violated.}
 \label{fig:f_vs_minfty_non_uniform}
\end{figure}

\subsection[Canonical ensemble]{Canonical ensemble: $F(\beta,L)$ (Fig.~\ref{fig:f_vs_minfty_non_uniform})}
\label{sec:free-energy}

The free energy $F(\beta,L)$ is also readily obtained from our numerical study. 
The results show that the dominance of phases in the canonical ensemble is the reverse of the microcanonical ensemble: in $d=4,5$ NUHPS are preferred over uniform strings at any temperature $T<T_H$, while in $d=6$ uniform strings are the dominant phase. As we have mentioned, corrections to the EFT are expected to modify the NUHPS for large enough non-uniformity but this must still be computed. Free strings only become relevant when the system is so close to the Hagedorn temperature that the large fluctuations render the HP model invalid, see \eqref{eq:qlimit}.

Again, these results confirm the conclusions of the heuristic comparison in Sec.~\ref{sec:basics}. They also resolve the issues with $F_b$ in $d=6$ mentioned there. 

\subsection{Stability criteria}\label{sec:stabcrit}

\paragraph{Dynamical stability.} Let us now examine whether, in light of these results, we can infer the existence of a dynamical GL instability for UHP strings. There is some inherent uncertainty in this analysis since the HP formalism is time-independent by nature, but general considerations can still be applied to this question.

First, we observe that Morse theory is indeed consistent with the assignments marked in our diagrams\footnote{We are grateful to Jinwei Chu for suggesting this kind of argument.}, depending on whether the UHPS is thinner or thicker than the critical string with zero-mode length,  $L\sqrt{\Delta_\beta} \gtrless \textsf{L}_*$. In $d=4,5$ we see that for masses smaller than $M_* = M(L_*)$ the symmetric phase (the UHPS) has a larger entropy than the symmetry-breaking phase (the NUHPS), implying that the latter is expected to be less stable. For $M>M_*$ only the symmetric phase remains. According to Morse theory, for $M>M_*$ the UHPS should possess one more unstable mode than for $M<M_*$. In the absence of any other instabilities, the UHPS will thus remain stable for small masses, $M<M_*$.\footnote{Let us note that \cite{Chen:2021dsw} found that HP balls in $d=3$ have negative Euclidean modes for fixed temperature, but not for fixed mass. However, the UHPS in $d=4$ can still have a GL zero mode for fixed mass, since its mass density can fluctuate while the total mass remains fixed.} The reason that UHPS become stable at small mass while black strings do so at large mass is that in both cases those are the regimes where they become fatter, see \eqref{eq:Mrb} and \eqref{eq:Mr0}.

In $d=6$ the situation is reversed. For low masses, $M<M_*$, when the UHPS is thinner, it has lower entropy than the NUHPS and thus is expected to be unstable. If the NUHPS is stable, then the UHPS for $M>M_*$ will also be expected to be stable---although both of them could be unstable to modes breaking other symmetries which are not captured by this analysis, a possibility that we will revisit later.

\paragraph{Local thermodynamic stability.} Let us now examine a different argument. It is a general property of uniform extended systems that the sign of their specific heat $C$ has a direct bearing on their dynamical stability. In the limit when the system is very long, its density fluctuations at long wavelengths behave as hydrodynamic sound waves, with an effective sound speed $v_s^2=s/C$, where $s$ is the entropy density. If $C<0$ these fluctuations do not propagate as sound but instead grow exponentially.\footnote{This argument \cite{Emparan:2012be} justifies the Gubser-Mitra correlated stability conjecture \cite{Gubser:2000mm} in the hydrodynamic regime.} For instance, the negative specific heat of the Schwarzschild solution directly implies a dynamical instability of black strings at very long wavelengths \cite{Emparan:2009at,Camps:2010br}. Since the uniform HP string in $d=4$ has $C_U<0$, a GL instability is expected at very long wavelengths,  consistent with the Morse theory analysis above. The UHPS in $d=5$ has infinite specific heat, so no definitive conclusion can be drawn. However, it seems reasonable to conjecture that, when higher-order corrections are included, its behavior will resemble that of $d=4$, with $C_U<0$ and a long-wavelength GL instability.

In addition to this, the sign of the specific heat has been argued in \cite{Reall:2001ag} to be correlated with the existence of a negative Euclidean mode for fixed-temperature fluctuations---which translates into the GL zero mode. While the original argument was made for black objects, it seems applicable to the HP system as well. Then, the fact that $C_U<0$ in $d=4$ aligns well with the existence of the GL zero mode, and also supports the expectation that corrections in $d=5$ will yield $C_U<0$, also in line with the expected GL instability at long wavelengths.

However, the arguments based on the sign of the specific heat become puzzling in $d=6$, where $C_U>0$ but a GL zero-mode (and a Euclidean negative mode) still exists, and moreover, Morse theory suggests an instability of the UHPS for wavelengths longer than the critical length. The resolution of this puzzle remains unclear, but we can offer some comments. The GL zero mode is a linearized mode at a finite wavenumber, specifically the critical $k$ found in Sec.~\ref{subsubsection:zeromode}. Its dispersion relation at very small $k$ need not exhibit hydrodynamic behavior; for instance, it could be a tachyonic mode with non-zero imaginary frequency at $k=0$. If this were the case, the sign of $C_U$ would have no bearing on long-wavelength stability. However, this does not resolve the apparent conflict between Morse theory and the arguments in \cite{Reall:2001ag}, since these concern the behavior at the critical wavenumber. One possible explanation is that there may be other modes not captured by the conventional GL analysis, which might lead to phases that could not be identified in our current analysis. Corrections to the EFT are unlikely to solve the problem since they should be small for the UHPS or slightly non-uniform solutions. Furthermore, it is important to remember that the HP framework is intrinsically time-independent, so the previous arguments, while plausible, may not be directly applicable in this context.

So far we have discussed only the stability of the uniform HP solutions, but Morse theory also suggests that for fixed mass, the NUHPS should be unstable in $d=4,5$ and stable in $d=6$. Conversely, when fixing the temperature, the NUHPS should be stable in $d=4,5$ and unstable in $d=6$. In all dimensions, the stability at fixed temperature (canonical ensemble) correlates well with the sign of the specific heat $C_{NU}$, but this relationship is reversed if we instead fix the mass (microcanonical ensemble). This differing behavior of the same solution in the two ensembles may be a reflection of the ensemble inequivalence away from the strict thermodynamic limit, which is likely broken by the self-gravitational corrections.

\subsection{Microcanonical HP and black hole phases} \label{subsec:allphases}

\begin{figure}
\begin{center}
    \includegraphics[width=0.48\textwidth]{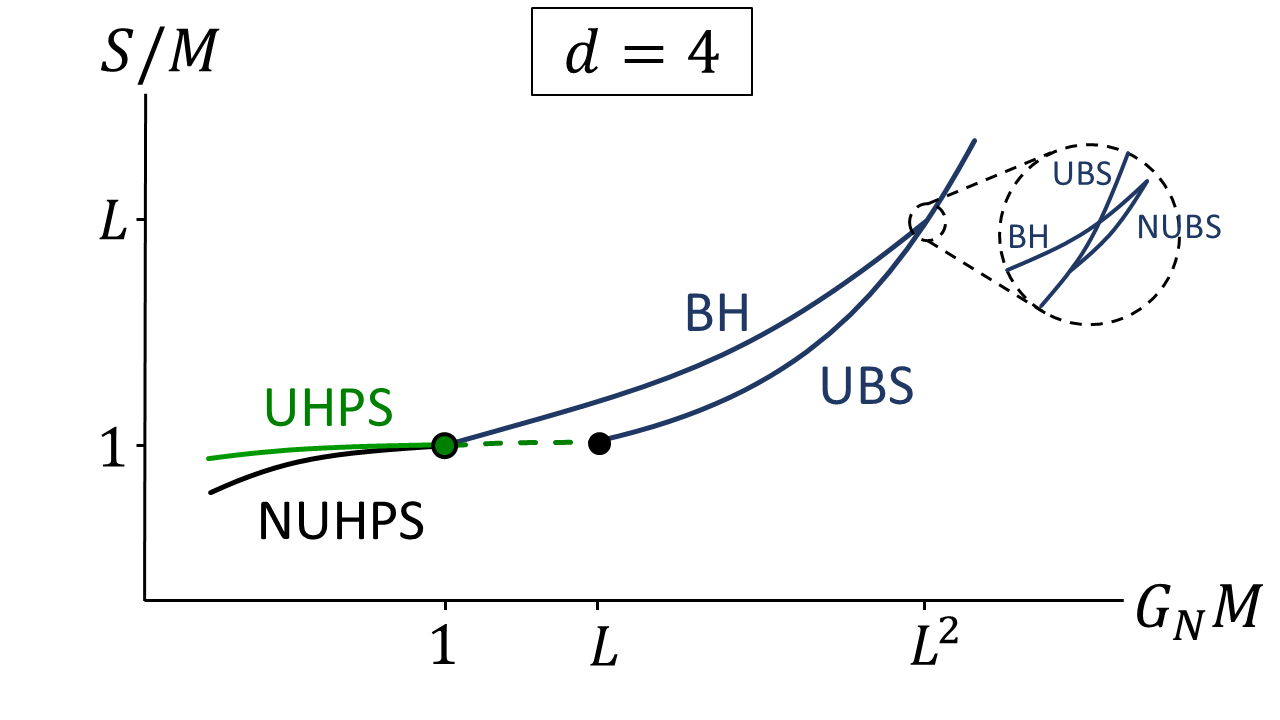}
    \includegraphics[width=0.48\textwidth]{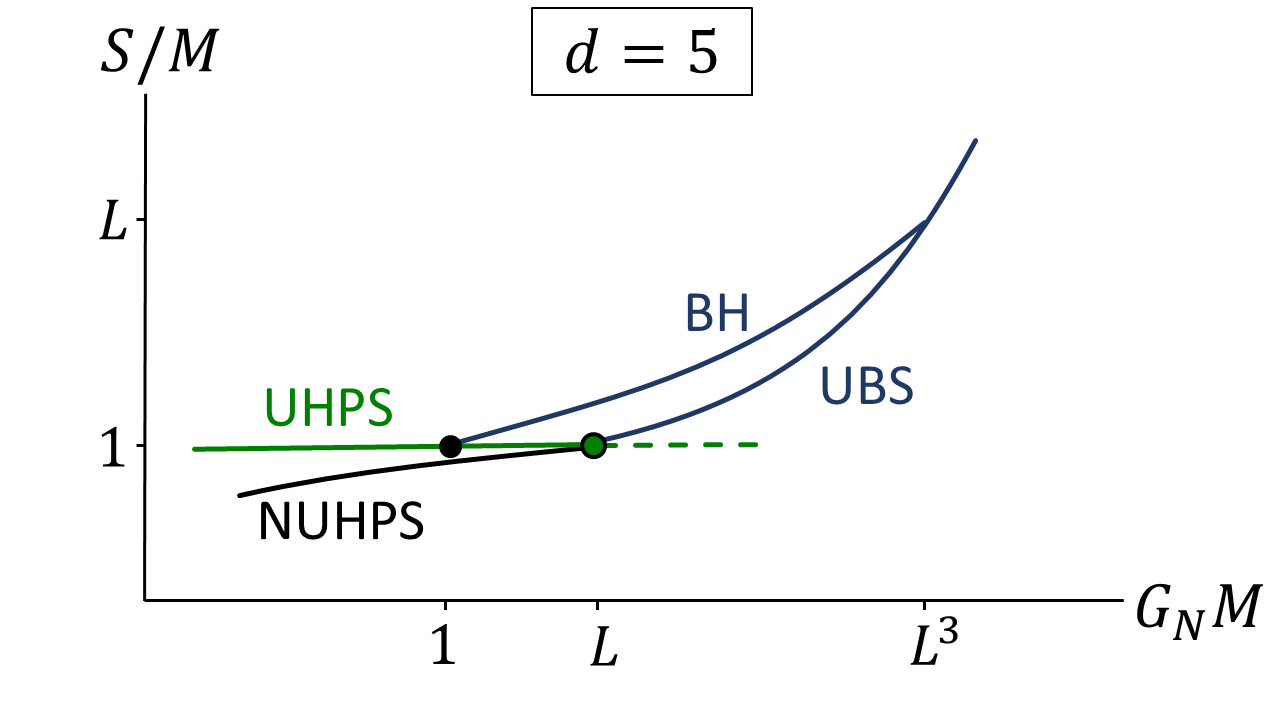}
    \includegraphics[width=0.48\textwidth]{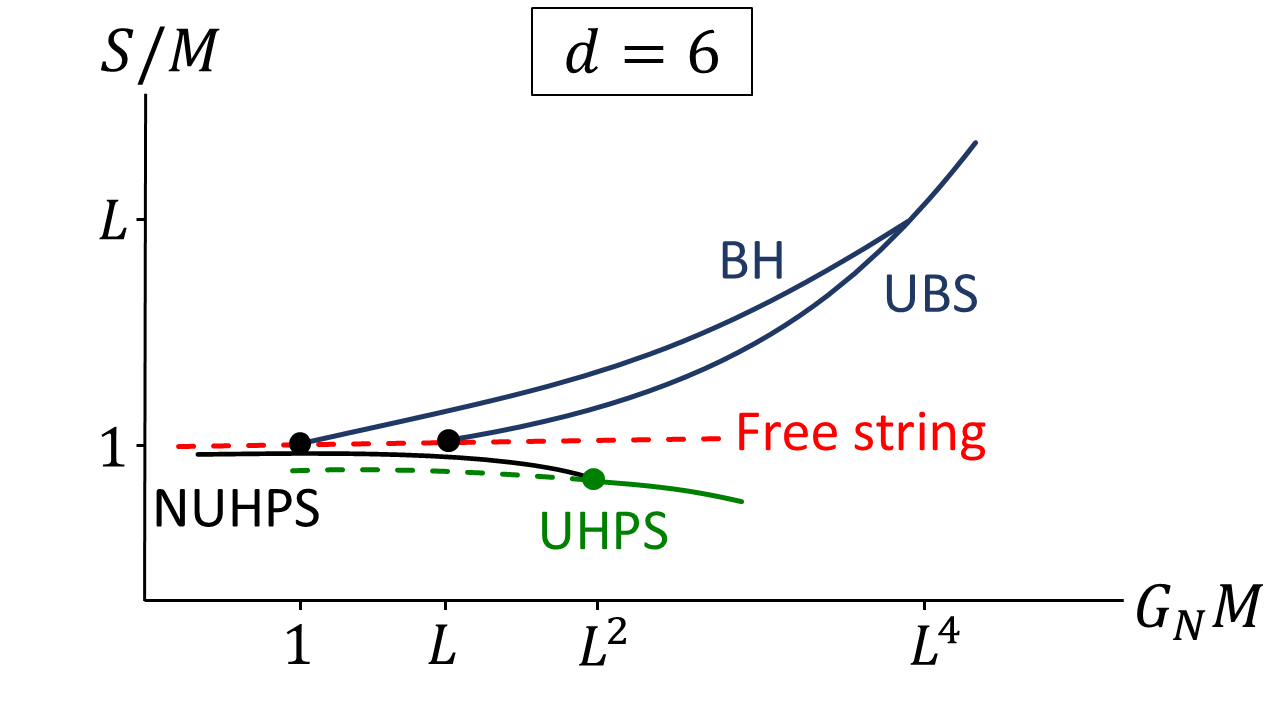}
\end{center}
\caption{\small Schematic phase diagrams including HP strings and black hole solutions: localized black holes (BH) and uniform black strings (UBS). For HP strings the information is essentially the same as in Fig.~\ref{fig:s_vs_m_non_uniform} but compressed around the Hagedorn line $S/M\sim 1$. The HP GL zero mode (green dot) occurs at $G_N M\sim L^{d-4}$. The correspondence transition  (black dots) for the BH occurs at $G_N M\sim 1$, and for the UBS at $G_N M\sim L$. The inset in the top-left figure zooms in on the merger at $G_N M\sim L^{d-2}$ between BH and UBS phases, involving also non-uniform black strings (NUBS). It does not play any role in our discussion, so we do not show it in $d=5,6$, although it is also present there.
}
  \label{fig:all_phases}
\end{figure}

Fig.~\ref{fig:all_phases} presents a schematic summary of our findings for the microcanonical HP phases combined with localized black holes (BH) and uniform black strings (UBS). Non-uniform black strings appear only at masses beyond the range relevant to our discussion, which are illustrated in the zoomed inset in Fig.~\ref{fig:all_phases}~(top-left).

The basic properties of black hole phases in a KK circle of length $L$ were discussed in Sec.~\ref{sec:basics}. Now we reintroduce $G_N$ while keeping string units, so $G_N\sim g^2$. At a given mass $M$, the entropies of BH and UBS are
\begin{equation}
    S_\textrm{BH} \sim M(G_N M)^{1/(d-2)}\,,\qquad  S_\textrm{UBS} \sim M\left(\frac{G_N M}{L}\right)^{1/(d-3)}\,.
\end{equation}
When comparing black holes and HP phases it is sufficient to consider the leading-order Hagedorn entropy of the latter, $S_{H}\sim M$.

We can now establish where the different phases---HP and black holes---meet. The merger between BH and UBS, parametrically close to the GL instability, happens when $S_\textrm{BH}=S_\textrm{UBS}$, i.e.,
\begin{equation}
    G_N M_{GL}\sim L^{d-2}\qquad \textrm{(UBS GL zero mode)}\,.
\end{equation}
For the UHPS, the GL instability occurs when the mass is such that $\textsf{L}=\textsf{L}_*\sim 1$. Using \eqref{MstarL}, this value is 
\begin{equation}
    G_N M_* \sim L^{d-4}\qquad \textrm{(UHP GL zero mode)}\,.
\end{equation}
The masses of the BH and UBS at the correspondence transition to string states, where their entropy equals the string entropy $S_{H}\sim M$, are
\begin{equation}\label{corrmass}
    G_N M_\textrm{BHc}\sim 1\,,\qquad  G_N M_\textrm{UBSc}\sim L\qquad \textrm{(Correspondence to strings)}\,.
\end{equation}

This information is illustrated in Fig.~\ref{fig:all_phases}. We observe that in $d=4$ the BH intersects with the UHPS at the mass where the latter becomes marginally stable. By contrast, in 
$d=5$ the BH transitions to a UHPS that is sufficiently light to be well within the stable regime. In $d=6$ the HP phases are always subdominant, and strings are only relevant, as almost free strings, for masses below the correspondence at $G_N M\sim 1$, i.e., $M/M_s \sim 1/g^2$.

The diagrams in Fig.~\ref{fig:all_phases} illustrate the likely fate of the UHPS with $M>M_*$ in $d=4,5$ as they undergo the GL instability. Rather than evolving into a non-uniform or localized HP phase---which does not exist at the same mass—--the increasing non-uniformity drives the UHPS to collapse into a black hole \cite{Chu:2024ggi}.

Finally, we note that our discussion has not addressed the sizes of the various phases. These sizes can be significant in transitions from strings to black holes in the dimensions where the string ball becomes very large, and heavier than \eqref{corrmass}, before collapsing into a black hole \cite{Horowitz:1997jc,Damour:1999aw}.  Our primary focus is instead on the scenarios where the initial state is a black hole phase, which will transition into a stringy phase (potentially puffing to larger size) once its mass decreases to the correspondence value \eqref{corrmass}.

\paragraph{Conclusion.} 
Our findings strongly suggest that uniform Horowitz-Polchinski (HP) string phases are classically stable in $d=4,5$  for masses below a critical threshold, $M<M_*$. For masses $M\lesssim M_s/g^2$ below the correspondence transition, stable uniform HP strings are the dominant microcanonical phases. Perhaps surprisingly, in this mass range there is no indication of a classical dynamical evolution leading to increasing non-uniformity. 
In $d=6$, uniform HP strings are expected to transition to a puffed-up almost-free string phase, likely bypassing the formation of stable non-uniform HP phases.

These results will provide the basis for our proposal of how strings deal with the endpoint of the GL instability.

\section{Stringy resolution of singular pinches}\label{sec:resolution}

We now apply these insights to the possible resolution in string theory of the singularities that arise when black strings become highly non-uniform, either in the space of static solutions or through the time evolution triggered by the GL instability.

\subsection{Pinch in static solutions}\label{sec:statpinch}

We have proven that as HP strings develop larger and larger non-uniformity, they smoothly go over into localized balls. This is a strong indication that string theory can smooth out the topology-changing transition between black strings and black holes in a KK circle. Can we use the HP formalism to resolve the singular pinch on the horizon?

This singularity is captured by an exact local model found in \cite{Kol:2002xz}. As the non-uniformity of static black string solutions grows, they develop a conical waist where the curvature grows arbitrarily large \cite{Kol:2003ja}. The local model is a self-similar Lorentzian conifold solution of the vacuum Einstein equations,
\begin{equation}
    ds^2=dr^2+\frac1{d-1}r^2\left( -\cos^2\psi\, dt^2+d\psi^2 \right)+  \frac{d-3}{d-1}r^2d\Omega_{d-2}\,,
\end{equation}
where $\psi\in [-\pi/2,\pi/2]$ and the endpoints of the interval correspond to the horizon. Near $r=0$ the curvature diverges as $(\textrm{Riemann})^2\sim 1/r^4$. One then expects that string theory is relevant when $r\lesssim \sqrt{\alpha'}$.

Unfortunately, even though this expectation is almost certainly true, the HP formalism does not seem to be directly applicable here. Since this geometry does not have any scale, one cannot take a limit where the curvature is everywhere small and the Newtonian approximation of the HP model uniformly applies. One might look for solutions of the HP equations that asymptote to a conical configuration, but the solutions that we have found at very high non-uniformity do not seem to exhibit it.

Therefore, when string theory takes over in the conical waist, it seems unlikely to be within the validity regime of the HP effective theory. The latter can nevertheless be used for black strings that reach the string scale before the conical waist develops. This requires that essentially all of the black string be near the string scale, and then it can be connected to a HP string. Depending on the number of dimensions and the ensemble that is considered, when the non-uniform black string transitions to a NUHPS, the dominant phase may change with a phase transition taking place. 

Let us mention that it might be interesting to study this problem in the large-$D$ limit, where the description of both the topology-changing transition and self-gravitating strings becomes significantly simpler \cite{Emparan:2019obu,Chen:2021emg} (see also \cite{Chu:2024ggi}).

\subsection{Dynamical pinch: The singularity is not near}\label{sec:notnear}

As discussed at the end of Sec.~\ref{sec:stabcrit}, if we start with a static uniform HP string with a mass below the correspondence transition, the Gregory-Laflamme instability disappears. Can this be relevant for the black string instability at late times, when the curvature along the thinning horizon reaches the string scale?

The classical evolution of the unstable black string in Einstein gravity has been studied in detail through numerical simulations in \cite{Lehner:2010pn,Figueras:2022zkg}. At late times, a thin neck forms, shrinking at a rate that is approximately linear in time. This rapid contraction suggests that when the stringy regime is reached, the time dependence may be too strong to rely on the static HP theory.

\paragraph{Stringy stalling.} Remarkably, there is evidence that the instability tapers off as the stringy stage is approached, with the evolution significantly slowing down. This is consistent with our string theory analysis, which suggests that when the string scale is reached, the black hole phase transitions into a stable uniform HP string in $d=4,5$, or into a string ball in $d=6$. This stabilization bootstraps the use of the HP solutions in the stringy regime.

This evidence comes from an impressive study of the deep non-linear evolution of the black string instability in the Einstein-Gauss-Bonnet (EGB) theory in five dimensions by Figueras, Kovács and Yao \cite{Figueras:2025}. This theory is not a proper truncation of the low-energy effective string theory to leading order in $\alpha'$, which in this context would at least require the introduction of the dilaton \cite{Callan:1988hs,Myers:1987qx}.\footnote{The axion and the two-form gauge field can be set to zero in our problem.} However, the dilaton is only sourced at quadratic order in $\alpha'$, so EGB may be a good proxy to string theory (plus, importantly, the theory admits a well-posed initial value formulation \cite{Papallo:2017qvl,Kovacs:2020ywu,Figueras:2024bba}). Of course, very close to the string scale, the higher-order corrections in the effective theory will also become important, meaning that one must change to a proper string theory description, and this is indeed where we hope our study is of use.

The starting point in \cite{Figueras:2025} is a slight perturbation of a uniform black string with a thickness much larger than $\sqrt{\alpha'}$ (where, in a slight abuse, $\alpha'$ denotes the EGB squared-length coupling). Without delving into the intricate details of the evolution that follows, what matters here is that, initially, the non-uniformity grows until a thin neck forms on the horizon, as expected. However, when the thickness approaches the string length $\sqrt{\alpha'}$, the process slows down and eventually halts, resulting in a long stretch of stable, thin, uniform black string. At this point, the effects of the GB term dominate, causing the effective theory to break down. What is relevant for our discussion is the strong deceleration of the evolution. Crucially, this slowdown occurs only when the sign of the GB coupling matches that of string theory; if the sign is reversed, the pinching continues to the bitter end. So far, the authors of \cite{Figueras:2025} have only performed evolutions in five dimensions within the framework of the EGB theory. It remains to be seen whether these results hold qualitatively in other dimensions or with other string-inspired corrections.



\paragraph{From GL to strings.} 
These results tentatively allow us to assemble the available evidence into a coherent picture. We interpret them as suggesting that once the system enters the stringy regime, the HP theory can reasonably serve as a guide for its subsequent evolution.

However, we must acknowledge the limitations of our current understanding. Our approach relies on comparing the dominance between thermodynamic phases, yet the nature of the near-stable segment of the thin black string identified in \cite{Figueras:2025} remains unclear. Should this segment be assimilated into the black hole (BH) or uniform black string (UBS) phases in Fig.~\ref{fig:all_phases}, or does it belong to neither? Our essential takeaway is the suggestion that, at least in $d=4,5$,
\emph{the black hole phase that is most stable as the string scale is reached, transitions into a stable uniform HP string.} Then, the quasi-stable segment of the pinching black string will transition into a stable UHPS.
In $d \geq 6$ it likely expands into a free-string ball.

The uncertainties in this picture leave room for more complex evolutions. For instance, the thinning black string might evolve into an unstable HP string. According to Fig.~\ref{fig:all_phases}, this would then collapse into a localized black hole, which at a later stage—--presumably through evaporation—--would reach the correspondence transition into a string ball. Intriguingly, in this scenario the topology-changing evolution from black string to localized black hole would be smoothed out through a classical evolution mediated by a stringy HP phase, without the formation of a singularity.

Another unknown is the length of the segment where the pinching black string reaches the string scale. It seems reasonable to assume that this segment would be longer, though not parametrically much longer, than the string-scale length. Notably, in $d=4$ the uniform HP string remains stable up to a length approximately $8$ times its thickness \eqref{L0HP}, which is itself larger than the string scale. Thus, a transition into an HP string segment thick enough to be stable appears plausible. However, one might also envision that the string-scale segment of the black string continues to grow, possibly devolving into the earlier scenario of an unstable HP string collapsing into localized black holes. 

While we cannot entirely rule out these more intricate evolutionary scenarios, they seem unnaturally contrived. The same mechanism that slows the pinching of the black string near the string scale is more likely to enhance, rather than diminish, the classical stability of the HP string. We hope that future progress into this problem will help refine the picture. For the time being, we adopt the simpler scenario outlined above as our primary framework for how the GL instability progresses into the stringy regime.

\paragraph{Non-singular endpoint of the GL instability.} Once it reaches this stringy stage, how does the system continue its evolution? Let us first consider the case of $d=4,5$. While classically stable, when the string coupling $g$ is non-zero the HP strings are expected to quantum-mechanically decay through radiation emission near the Hagedorn temperature. They will continue to evaporate in this way until their mass reaches the lower bound in  \eqref{eq:Mbound}, at which point their self-gravity becomes negligible and they are better described as free string balls. These free string balls will continue to radiate at the Hagedorn temperature \cite{Amati:1999fv}, until the evaporation process is complete. At this point, the black string finally fragments into separate pieces, bringing the GL evolution to a close.

How long does the HP phase take to evaporate? While no concrete calculations are available (and it is unclear how to perform them), it seems reasonable to expect that, similarly to Hawking emission in black holes, the HP ball radiates at a rate of one quantum of energy $\sim M_s$ in a string time $M_s^{-1}$. To reach the final free-string phase, at least an $O(1)$ fraction of the initial energy must be radiated by the HP ball, and this will take a time $\propto  S M_s^{-1}\gg  M_s^{-1}$. It is unclear whether the evaporation of the free string will continue at the same rate or faster, but the important point is that the entire stringy phase of the GL instability will last a time much longer than if the pinching had continued at the classical evolution rate at the moment of the transition into the stringy regime, $\Gamma \propto M_s$.

The duration of the evaporation stage in $d \geq 6$, where free-string balls dominate the stringy phase of the evolution, is somewhat more uncertain. However, for small coupling $g$, the evaporation is likely to still be much slower than the classical shrinking rate, possibly taking a time of the order $\sim S M_s^{-1}$, similar to the behavior in $d=4,5$.

\medskip

This concludes our proposed picture of the GL instability up until string theory brings about the fragmentation of the black string. In the subsequent evolution, possibly analogous to the behavior of fluid jets \cite{Eggers:1997zz}, classical gravity will resume governing the dynamics. This is left for future work.

\section{Outlook}
\label{sec:conclusion}

Our study of the HP phases and their thermodynamics admits several extensions, e.g., to strings with electric NS charges, such as an F-string charge along the circle direction. This is expected to further enhance the stability of the uniform solution. The situation in $d\geq 6$, where corrections to the HP effective theory  become important, also deserves closer attention.

The most exciting prospects lie in advancing our understanding of how string theory handles the best-attested violations of weak Cosmic Censorship. 
Admittedly, our conclusions in Sec.~\ref{sec:notnear} about the stringy resolution of the GL naked singularities are not based on a rigorous calculation of the evolution, which seems far beyond current capabilities. Instead, they synthesize a substantial body of evidence to construct a coherent picture. One way to strengthen the evidence is through a more detailed and extensive investigation of the role of $\alpha'$ effects in the late stages of the GL instability. 

Our proposal that string theory dampens the approach to a naked singularity, to then resolve it in a slower evaporation phase, should also be explored in the context of critical collapse \cite{Choptuik:1992jv}. More generally, the resolution of the static pinches, where our progress using the HP theory was limited, will be helped by studying $\alpha'$ corrections in the effective gravitational theory. Work in these directions is currently underway.

\section*{Acknowledgments}

We thank Pablo Cano and Rob Myers for useful conversations, and especially Pau Figueras for sharing with us the results of his ongoing work with Kovács and Yao in \cite{Figueras:2025}. We are also very grateful to Jinwei Chu for his comments on the first version of this article and for sharing with us other insights into this problem.
RE is supported by MICINN grant PID2022-136224NB-C22, AGAUR grant 2021 SGR 00872, and State
Research Agency of MICINN through the ``Unit of Excellence María de Maeztu 2020-2023'' award to the Institute of Cosmos Sciences (CEX2019-000918-M).
The work of MSG was partially supported by  the European Research Council (ERC) under the European Union’s Horizon 2020 research and innovation program (grant agreement No758759).
MT is supported by the European Research Council (ERC) under the European Union’s
Horizon 2020 research and innovation programme (grant agreement No 852386). MT is also supported by the Emmy Noether Fellowship program at the Perimeter Institute for Theoretical Physics.

\bibliography{HPbib}

\providecommand{\href}[2]{#2}\begingroup\raggedright\begin{thebibliography}{10}

\bibitem{Gregory:1993vy}
R.~Gregory and R.~Laflamme, \emph{{Black strings and p-branes are unstable}},
  \href{https://doi.org/10.1103/PhysRevLett.70.2837}{\emph{Phys. Rev. Lett.}
  {\bfseries 70} (1993) 2837}
  [\href{https://arxiv.org/abs/hep-th/9301052}{{\ttfamily hep-th/9301052}}].

\bibitem{Gregory:1994bj}
R.~Gregory and R.~Laflamme, \emph{{The instability of charged black strings and
  p-branes}}, \href{https://doi.org/10.1016/0550-3213(94)90206-2}{\emph{Nucl.
  Phys. B} {\bfseries 428} (1994) 399}
  [\href{https://arxiv.org/abs/hep-th/9404071}{{\ttfamily hep-th/9404071}}].

\bibitem{Susskind:1993ws}
L.~Susskind, \emph{{Some speculations about black hole entropy in string
  theory}},  \href{https://arxiv.org/abs/hep-th/9309145}{{\ttfamily
  hep-th/9309145}}.

\bibitem{Horowitz:1996nw}
G.~T. Horowitz and J.~Polchinski, \emph{{A correspondence principle for black
  holes and strings}},
  \href{https://doi.org/10.1103/PhysRevD.55.6189}{\emph{Phys. Rev.} {\bfseries
  D55} (1997) 6189} [\href{https://arxiv.org/abs/hep-th/9612146}{{\ttfamily
  hep-th/9612146}}].

\bibitem{Horowitz:1997jc}
G.~T. Horowitz and J.~Polchinski, \emph{{Selfgravitating fundamental strings}},
  \href{https://doi.org/10.1103/PhysRevD.57.2557}{\emph{Phys. Rev. D}
  {\bfseries 57} (1998) 2557}
  [\href{https://arxiv.org/abs/hep-th/9707170}{{\ttfamily hep-th/9707170}}].

\bibitem{Lehner:2010pn}
L.~Lehner and F.~Pretorius, \emph{{Black Strings, Low Viscosity Fluids, and
  Violation of Cosmic Censorship}},
  \href{https://doi.org/10.1103/PhysRevLett.105.101102}{\emph{Phys. Rev. Lett.}
  {\bfseries 105} (2010) 101102}
  [\href{https://arxiv.org/abs/1006.5960}{{\ttfamily 1006.5960}}].

\bibitem{Figueras:2022zkg}
P.~Figueras, T.~Fran\c{c}a, C.~Gu and T.~Andrade, \emph{{Endpoint of the
  Gregory-Laflamme instability of black strings revisited}},
  \href{https://doi.org/10.1103/PhysRevD.107.044028}{\emph{Phys. Rev. D}
  {\bfseries 107} (2023) 044028}
  [\href{https://arxiv.org/abs/2210.13501}{{\ttfamily 2210.13501}}].

\bibitem{Wiseman:2002zc}
T.~Wiseman, \emph{{Static axisymmetric vacuum solutions and nonuniform black
  strings}}, \href{https://doi.org/10.1088/0264-9381/20/6/308}{\emph{Class.
  Quant. Grav.} {\bfseries 20} (2003) 1137}
  [\href{https://arxiv.org/abs/hep-th/0209051}{{\ttfamily hep-th/0209051}}].

\bibitem{Kleihaus:2006ee}
B.~Kleihaus, J.~Kunz and E.~Radu, \emph{{New nonuniform black string
  solutions}}, \href{https://doi.org/10.1088/1126-6708/2006/06/016}{\emph{JHEP}
  {\bfseries 06} (2006) 016}
  [\href{https://arxiv.org/abs/hep-th/0603119}{{\ttfamily hep-th/0603119}}].

\bibitem{Sorkin:2006wp}
E.~Sorkin, \emph{{Non-uniform black strings in various dimensions}},
  \href{https://doi.org/10.1103/PhysRevD.74.104027}{\emph{Phys. Rev. D}
  {\bfseries 74} (2006) 104027}
  [\href{https://arxiv.org/abs/gr-qc/0608115}{{\ttfamily gr-qc/0608115}}].

\bibitem{Harmark:2007md}
T.~Harmark, V.~Niarchos and N.~A. Obers, \emph{{Instabilities of black strings
  and branes}}, \href{https://doi.org/10.1088/0264-9381/24/8/R01}{\emph{Class.
  Quant. Grav.} {\bfseries 24} (2007) R1}
  [\href{https://arxiv.org/abs/hep-th/0701022}{{\ttfamily hep-th/0701022}}].

\bibitem{Figueras:2012xj}
P.~Figueras, K.~Murata and H.~S. Reall, \emph{{Stable non-uniform black strings
  below the critical dimension}},
  \href{https://doi.org/10.1007/JHEP11(2012)071}{\emph{JHEP} {\bfseries 11}
  (2012) 071} [\href{https://arxiv.org/abs/1209.1981}{{\ttfamily 1209.1981}}].

\bibitem{Kalisch:2016fkm}
M.~Kalisch and M.~Ansorg, \emph{{Pseudo-spectral construction of non-uniform
  black string solutions in five and six spacetime dimensions}},
  \href{https://doi.org/10.1088/0264-9381/33/21/215005}{\emph{Class. Quant.
  Grav.} {\bfseries 33} (2016) 215005}
  [\href{https://arxiv.org/abs/1607.03099}{{\ttfamily 1607.03099}}].

\bibitem{Kalisch:2017bin}
M.~Kalisch, S.~M\"ockel and M.~Ammon, \emph{{Critical behavior of the black
  hole/black string transition}},
  \href{https://doi.org/10.1007/JHEP08(2017)049}{\emph{JHEP} {\bfseries 08}
  (2017) 049} [\href{https://arxiv.org/abs/1706.02323}{{\ttfamily
  1706.02323}}].

\bibitem{Cardona:2018shd}
B.~Cardona and P.~Figueras, \emph{{Critical Kaluza-Klein black holes and black
  strings in D = 10}},
  \href{https://doi.org/10.1007/JHEP11(2018)120}{\emph{JHEP} {\bfseries 11}
  (2018) 120} [\href{https://arxiv.org/abs/1806.11129}{{\ttfamily
  1806.11129}}].

\bibitem{Kol:2002xz}
B.~Kol, \emph{{Topology change in general relativity, and the black hole black
  string transition}},
  \href{https://doi.org/10.1088/1126-6708/2005/10/049}{\emph{JHEP} {\bfseries
  10} (2005) 049} [\href{https://arxiv.org/abs/hep-th/0206220}{{\ttfamily
  hep-th/0206220}}].

\bibitem{Kol:2003ja}
B.~Kol and T.~Wiseman, \emph{{Evidence that highly nonuniform black strings
  have a conical waist}},
  \href{https://doi.org/10.1088/0264-9381/20/15/315}{\emph{Class. Quant. Grav.}
  {\bfseries 20} (2003) 3493}
  [\href{https://arxiv.org/abs/hep-th/0304070}{{\ttfamily hep-th/0304070}}].

\bibitem{Friedberg:1986tp}
R.~Friedberg, T.~D. Lee and Y.~Pang, \emph{{Mini-soliton stars}},
  \href{https://doi.org/10.1103/PhysRevD.35.3640}{\emph{Phys. Rev. D}
  {\bfseries 35} (1987) 3640}.

\bibitem{Jones:1995wb}
K.~R.~W. Jones, \emph{{Newtonian quantum gravity}},
  \href{https://doi.org/10.1071/PH951055}{\emph{Austral. J. Phys.} {\bfseries
  48} (1995) 1055} [\href{https://arxiv.org/abs/quant-ph/9507001}{{\ttfamily
  quant-ph/9507001}}].

\bibitem{Chen:2021dsw}
Y.~Chen, J.~Maldacena and E.~Witten, \emph{{On the black hole/string
  transition}}, \href{https://doi.org/10.1007/JHEP01(2023)103}{\emph{JHEP}
  {\bfseries 01} (2023) 103}
  [\href{https://arxiv.org/abs/2109.08563}{{\ttfamily 2109.08563}}].

\bibitem{Gregory:1987nb}
R.~Gregory and R.~Laflamme, \emph{{Hypercylindrical black holes}},
  \href{https://doi.org/10.1103/PhysRevD.37.305}{\emph{Phys. Rev. D} {\bfseries
  37} (1988) 305}.

\bibitem{Reall:2001ag}
H.~S. Reall, \emph{{Classical and thermodynamic stability of black branes}},
  \href{https://doi.org/10.1103/PhysRevD.64.044005}{\emph{Phys. Rev. D}
  {\bfseries 64} (2001) 044005}
  [\href{https://arxiv.org/abs/hep-th/0104071}{{\ttfamily hep-th/0104071}}].

\bibitem{Balthazar:2022szl}
B.~Balthazar, J.~Chu and D.~Kutasov, \emph{{Winding Tachyons and Stringy Black
  Holes}},  \href{https://arxiv.org/abs/2204.00012}{{\ttfamily 2204.00012}}.

\bibitem{Balthazar:2022hno}
B.~Balthazar, J.~Chu and D.~Kutasov, \emph{{On small black holes in string
  theory}}, \href{https://doi.org/10.1007/JHEP03(2024)116}{\emph{JHEP}
  {\bfseries 03} (2024) 116}
  [\href{https://arxiv.org/abs/2210.12033}{{\ttfamily 2210.12033}}].

\bibitem{Figueras:2025}
P.~Figueras, A.~D. Kovács and S.~Yao, ``{The evolution of the Gregory-Laflamme
  instability of five-dimensional black strings in Einstein-Gauss-Bonnet
  gravity}.'' {In progress}.

\bibitem{Chu:2024ggi}
J.~Chu, \emph{{From Black Strings to Fundamental Strings: Non-uniformity and
  Phase Transitions}},  \href{https://arxiv.org/abs/2410.23597}{{\ttfamily
  2410.23597}}.

\bibitem{Emparan:2018bmi}
R.~Emparan, R.~Luna, M.~Mart\'\i{}nez, R.~Suzuki and K.~Tanabe, \emph{{Phases
  and Stability of Non-Uniform Black Strings}},
  \href{https://doi.org/10.1007/JHEP05(2018)104}{\emph{JHEP} {\bfseries 05}
  (2018) 104} [\href{https://arxiv.org/abs/1802.08191}{{\ttfamily
  1802.08191}}].

\bibitem{Polchinski:1985zf}
J.~Polchinski, \emph{{Evaluation of the One Loop String Path Integral}},
  \href{https://doi.org/10.1007/BF01210791}{\emph{Commun. Math. Phys.}
  {\bfseries 104} (1986) 37}.

\bibitem{Atick:1988si}
J.~J. Atick and E.~Witten, \emph{{The Hagedorn Transition and the Number of
  Degrees of Freedom of String Theory}},
  \href{https://doi.org/10.1016/0550-3213(88)90151-4}{\emph{Nucl. Phys. B}
  {\bfseries 310} (1988) 291}.

\bibitem{Horowitz:1996cj}
G.~T. Horowitz and D.~Marolf, \emph{{Counting states of black strings with
  traveling waves. II}},
  \href{https://doi.org/10.1103/PhysRevD.55.846}{\emph{Phys. Rev.} {\bfseries
  D55} (1997) 846} [\href{https://arxiv.org/abs/hep-th/9606113}{{\ttfamily
  hep-th/9606113}}].

\bibitem{Brustein:2021ifl}
R.~Brustein and Y.~Zigdon, \emph{{Effective field theory for closed strings
  near the Hagedorn temperature}},
  \href{https://doi.org/10.1007/JHEP04(2021)107}{\emph{JHEP} {\bfseries 04}
  (2021) 107} [\href{https://arxiv.org/abs/2101.07836}{{\ttfamily
  2101.07836}}].

\bibitem{Albertini:2024hwi}
E.~Albertini, D.~Platt and T.~Wiseman, \emph{{Towards a uniqueness theorem for
  static black holes in Kaluza-Klein theory with small circle size}},
  \href{https://arxiv.org/abs/2410.20967}{{\ttfamily 2410.20967}}.

\bibitem{Dias:2015nua}
O.~J.~C. Dias, J.~E. Santos and B.~Way, \emph{{Numerical Methods for Finding
  Stationary Gravitational Solutions}},
  \href{https://doi.org/10.1088/0264-9381/33/13/133001}{\emph{Class. Quant.
  Grav.} {\bfseries 33} (2016) 133001}
  [\href{https://arxiv.org/abs/1510.02804}{{\ttfamily 1510.02804}}].

\bibitem{Ceplak:2023afb}
N.~\v{C}eplak, R.~Emparan, A.~Puhm and M.~Toma\v{s}evi\'c, \emph{{The
  correspondence between rotating black holes and fundamental strings}},
  \href{https://doi.org/10.1007/JHEP11(2023)226}{\emph{JHEP} {\bfseries 11}
  (2023) 226} [\href{https://arxiv.org/abs/2307.03573}{{\ttfamily
  2307.03573}}].

\bibitem{Emparan:2012be}
R.~Emparan and M.~Martinez, \emph{{Black Branes in a Box: Hydrodynamics,
  Stability, and Criticality}},
  \href{https://doi.org/10.1007/JHEP07(2012)120}{\emph{JHEP} {\bfseries 07}
  (2012) 120} [\href{https://arxiv.org/abs/1205.5646}{{\ttfamily 1205.5646}}].

\bibitem{Gubser:2000mm}
S.~S. Gubser and I.~Mitra, \emph{{The Evolution of unstable black holes in
  anti-de Sitter space}},
  \href{https://doi.org/10.1088/1126-6708/2001/08/018}{\emph{JHEP} {\bfseries
  08} (2001) 018} [\href{https://arxiv.org/abs/hep-th/0011127}{{\ttfamily
  hep-th/0011127}}].

\bibitem{Emparan:2009at}
R.~Emparan, T.~Harmark, V.~Niarchos and N.~A. Obers, \emph{{Essentials of
  Blackfold Dynamics}},
  \href{https://doi.org/10.1007/JHEP03(2010)063}{\emph{JHEP} {\bfseries 03}
  (2010) 063} [\href{https://arxiv.org/abs/0910.1601}{{\ttfamily 0910.1601}}].

\bibitem{Camps:2010br}
J.~Camps, R.~Emparan and N.~Haddad, \emph{{Black Brane Viscosity and the
  Gregory-Laflamme Instability}},
  \href{https://doi.org/10.1007/JHEP05(2010)042}{\emph{JHEP} {\bfseries 05}
  (2010) 042} [\href{https://arxiv.org/abs/1003.3636}{{\ttfamily 1003.3636}}].

\bibitem{Damour:1999aw}
T.~Damour and G.~Veneziano, \emph{{Selfgravitating fundamental strings and
  black holes}},
  \href{https://doi.org/10.1016/S0550-3213(99)00596-9}{\emph{Nucl. Phys.}
  {\bfseries B568} (2000) 93}
  [\href{https://arxiv.org/abs/hep-th/9907030}{{\ttfamily hep-th/9907030}}].

\bibitem{Emparan:2019obu}
R.~Emparan and R.~Suzuki, \emph{{Topology-changing horizons at large D as Ricci
  flows}}, \href{https://doi.org/10.1007/JHEP07(2019)094}{\emph{JHEP}
  {\bfseries 07} (2019) 094}
  [\href{https://arxiv.org/abs/1905.01062}{{\ttfamily 1905.01062}}].

\bibitem{Chen:2021emg}
Y.~Chen and J.~Maldacena, \emph{{String scale black holes at large D}},
  \href{https://doi.org/10.1007/JHEP01(2022)095}{\emph{JHEP} {\bfseries 01}
  (2022) 095} [\href{https://arxiv.org/abs/2106.02169}{{\ttfamily
  2106.02169}}].

\bibitem{Callan:1988hs}
C.~G. Callan, Jr., R.~C. Myers and M.~J. Perry, \emph{{Black Holes in String
  Theory}}, \href{https://doi.org/10.1016/0550-3213(89)90172-7}{\emph{Nucl.
  Phys. B} {\bfseries 311} (1989) 673}.

\bibitem{Myers:1987qx}
R.~C. Myers, \emph{{Superstring Gravity and Black Holes}},
  \href{https://doi.org/10.1016/0550-3213(87)90402-0}{\emph{Nucl. Phys. B}
  {\bfseries 289} (1987) 701}.

\bibitem{Papallo:2017qvl}
G.~Papallo and H.~S. Reall, \emph{{On the local well-posedness of Lovelock and
  Horndeski theories}},
  \href{https://doi.org/10.1103/PhysRevD.96.044019}{\emph{Phys. Rev. D}
  {\bfseries 96} (2017) 044019}
  [\href{https://arxiv.org/abs/1705.04370}{{\ttfamily 1705.04370}}].

\bibitem{Kovacs:2020ywu}
A.~D. Kov\'acs and H.~S. Reall, \emph{{Well-posed formulation of Lovelock and
  Horndeski theories}},
  \href{https://doi.org/10.1103/PhysRevD.101.124003}{\emph{Phys. Rev. D}
  {\bfseries 101} (2020) 124003}
  [\href{https://arxiv.org/abs/2003.08398}{{\ttfamily 2003.08398}}].

\bibitem{Figueras:2024bba}
P.~Figueras, A.~Held and A.~D. Kov\'acs, \emph{{Well-posed initial value
  formulation of general effective field theories of gravity}},
  \href{https://arxiv.org/abs/2407.08775}{{\ttfamily 2407.08775}}.

\bibitem{Amati:1999fv}
D.~Amati and J.~G. Russo, \emph{{Fundamental strings as black bodies}},
  \href{https://doi.org/10.1016/S0370-2693(99)00375-5}{\emph{Phys. Lett. B}
  {\bfseries 454} (1999) 207}
  [\href{https://arxiv.org/abs/hep-th/9901092}{{\ttfamily hep-th/9901092}}].

\bibitem{Eggers:1997zz}
J.~Eggers, \emph{{Nonlinear dynamics and breakup of free-surface flows}},
  \href{https://doi.org/10.1103/RevModPhys.69.865}{\emph{Rev. Mod. Phys.}
  {\bfseries 69} (1997) 865}.

\bibitem{Choptuik:1992jv}
M.~W. Choptuik, \emph{{Universality and scaling in gravitational collapse of a
  massless scalar field}},
  \href{https://doi.org/10.1103/PhysRevLett.70.9}{\emph{Phys. Rev. Lett.}
  {\bfseries 70} (1993) 9}.

\end{thebibliography}\endgroup

\end{document}